\documentclass[aps,pra,twocolumn,superscriptaddress,footinbib,showpacs]{revtex4-1}
\usepackage{graphicx}
\usepackage{color,ulem}

\renewcommand{\emph}[1]{{\it #1}}

\begin{document}

\title{Unsharp continuous measurement of a Bose-Einstein condensate: \\ full quantum state estimation and the transition to classicality}

\author{Moritz Hiller}
\affiliation{Physikalisches Institut, Albert-Ludwigs-Universit\"at Freiburg, Hermann-Herder-Str. 3, 79104 Freiburg, Germany}
\affiliation{Institute for Theoretical Physics, Vienna University of Technology, Wiedner Hauptstra{\ss}e 8-10/136, 1040 Vienna, Austria}
\author{Magnus Rehn}
\affiliation{School of Chemistry and Physics, University of KwaZulu-Natal, Durban, South Africa} 
\affiliation{National Institute of Theoretical Physics (Durban Node), South Africa}
\author{Francesco Petruccione}
\affiliation{School of Chemistry and Physics, University of KwaZulu-Natal, Durban, South Africa} 
\affiliation{National Institute of Theoretical Physics (Durban Node), South Africa}
\author{Andreas Buchleitner}
\affiliation{Physikalisches Institut, Albert-Ludwigs-Universit\"at Freiburg, Hermann-Herder-Str. 3, 79104 Freiburg, Germany}
\author{Thomas Konrad} 
\affiliation{School of Chemistry and Physics, University of KwaZulu-Natal, Durban, South Africa} 
\affiliation{National Institute of Theoretical Physics (Durban Node), South Africa}

\begin{abstract}
We study a Bose-Einstein condensate (BEC) in a double-well potential subject to an unsharp continuous measurement of the atom number in one of the two wells.
We investigate the back action of the measurement on the quantum dynamics and the viability to monitor the ensuing time evolution.
For vanishing inter-atomic interactions, mainly the expectation values of the measured local observable can be inferred from the measurement record.
Conversely, in the presence of moderate inter-atomic interactions, the entire many-body state  --modified by the measurement--  is monitored with unit fidelity and, at the same time, the measurement effects a transition from quantum to mean-field (classical) behavior of the BEC.
We show that this perfect state estimation is possible because the inter-atomic interactions enhance the information gained via the measurement.
\end{abstract}
 

\maketitle

\section{Introduction}

Continuous measurements enable the monitoring and control of {\it individual} quantum systems in real time, a prerogative in applications of quantum mechanics such as quantum information processing and high-precision measurements \cite{PlenioKnight98, JacobsSteck06, WisemanMilburn09, Barchielli09}. 
Immanent to quantum physics, the trade-off between information gain and disturbance of the system requires  the continuous measurement to be of low eigenvalue resolution per time unit in order to limit the measurement-induced alteration of the dynamics.
Such an {\it unsharp} continuous measurement can be realized, e.g., by consecutive indirect measurements: the system interacts in a rapid succession with a sequence of quantum probes which are subsequently measured.
The strength of the continuous measurement can then be controlled via the width of the probe's wave function or by  tuning the interaction with the system.
Continuous measurements based on sequences of indirect measurements have been designed, e.g., to monitor single observables like the position of a quantum particle \cite{CavesMilburn87}, the photon number \cite{AudretschKonradScherer02} in cavity QED, the charge distribution in coupled quantum dots \cite{Korotkov00a}, and can be used to track Rabi oscillations  \cite{AudretschKonradScherer01, AudretschKleeKonrad04}.
In further applications, unsharp measurements have been employed to estimate the pre-measurement state of an ensemble of identically prepared systems \cite{Silberfarb.et.al.05, Smith.et.al.06}, to determine the frequency of Rabi oscillations \cite {ChaseGeremia09}, and, combined with feedback loops, to cool atoms \cite{Steck.et.al.06} and steer the system into a targeted state \cite{ShabaniJacobs08}.

Recently, ultracold atoms in trapping potentials were brought into focus of unsharp measurements as they represent interacting many-body systems under exquisite experimental control.
Among the observations are, e.g., the establishment of macroscopic coherence \cite{RW97,CorneyMilburn98,DDO02}, even in the presence of environmental decoherence  \cite{RW98}, the detection and preparation of Mott- or superfluid states \cite{MMR07,MR09}, the enhancement of macroscopic self-trapping \cite{LL06},  and the possibility of feedback control \cite{SHC10}.

In the present contribution, we study a continuously measured Bose-Einstein condensate (BEC) in a double-well potential.
Our motivation is driven by two key-targets:
On the one hand, we strive to explain the influence of the measurement on the quantum many-body dynamics.
It was shown that the dephasing between the BECs in the two wells -- which arises due to the inter-atomic interactions \cite{MCWW97,VA00,ZNGO10} -- can be reduced by the measurement \cite{RW97,RW98,CorneyMilburn98}.
We unveil the underlying mechanism and show that the observed behavior is the manifestation of a transition from quantum to classical (mean-field) dynamics of the BEC.

On the other hand, we address the question whether the measurement of a {\it local} observable of a many-body system can yield {\it complete knowledge} of its wave function.
If so, this would allow complete control of the system via feedback depending on the measurement record.
To tackle this problem, we employ a versatile technique \cite{Diosi.et.al06} which was successfully applied to estimate the state of a {\it single} particle in several one- and two-dimensional potentials (among them the classically chaotic Henon-Heiles potential) \cite{Konrad.et.al10}, as well as of a two-level system in the presence of noise \cite{KonradUys12}.
We will show, that the strength of the inter-atomic interactions in the BEC decides whether or not such a state estimation is possible.

The narrative of this work is as follows:
We first introduce the Bose-Hubbard Hamiltonian in angular momentum representation as the model for a BEC in a double-well potential.
In Section \ref{sec.homodyne}, the realization of the continuous measurement of the atom number according to Corney and Milburn \cite{CorneyMilburn98} is discussed together with the stochastic equations that describe the coupled dynamics of state and measurement record.
The procedure of state estimation is laid out in Section \ref{sec.estimate} while Sections \ref{sec.results_u0}-\ref{sec.results_u1} contains our results:
for moderate measurement strengths, we analyze quantum dynamics and estimation fidelity for the cases of vanishing and weak inter-atomic interactions.

\section{Model: BEC in a double-well potential}
A Bose-Einstein condensate of $N$ ultra-cold bosons in a double-well potential is described by the Bose-Hubbard (BH) Hamiltonian \cite{MCWW97,JBCGZ98}:
\begin{equation}
\label{bosehubb}
\hat{H}_{BH} =  U\hat{J}_z^2 - K\hat{J_x},
\end{equation} 
with angular momentum operators  $\hat{J}_x = (\hat{b}_1^{\dagger}\hat{b}_2 + \hat{b}_1\hat{b}_2^{\dagger})/2$,
$\hat{J}_y =  i (\hat{b}_2^{\dagger}\hat{b}_1-\hat{b}_1^{\dagger}\hat{b}_2)/2$, $\hat{J}_z =   (\hat{b}_1^{\dagger}\hat{b}_1-\hat{b}_2^{\dagger}\hat{b}_2)/2$, and $\hat J_+=\hat{b}_1^{\dagger}\hat{b}_2$.
Here, $\hat{b}_i^{(\dagger)}$ are the bosonic annihilation (creation) operators, and $\hat{n}_i=\hat{b}_i^\dagger\hat{b}_i$ is the number counting operator at site $i$
\footnote{\label{foot.bias} Numerically, we add a small bias of the order of $10^{-2}\protect\hat n_1$, in order to avoid unstable macroscopic superposition states \cite{DHC07} that result in numerical artifacts.}.
In (\ref{bosehubb}), $U$ and $K$ parametrize the on-site inter-atomic interaction and the tunnelling strength, respectively. 
Experimentally, both parameters can be independently controlled via the height of the potential barrier and by additional magnetic fields that induce Feshbach resonances \cite{MO06,CGJT10}. 
Apart from the total energy $E$, also the total particle number $n_1+n_2=N=2j$ is a constant of motion since $[\hat{J}^2, \hat{H}_{BH}]=0$.
Thus, the states of the Fock number basis, i.e., the eigenstates of $\hat J_z$, can be equally expressed as $|n_1,n_2\rangle=|n_1,N-n_1\rangle=|m\rangle$, where the angular momentum $m=n_1-N/2$ ranges from $-j$ to $j=N/2$.
The physical interpretation of the operators $\hat{J}_i$ is as follows: $\hat{J}_z$ measures the particle imbalance between the
wells, while $\hat{J}_y$ represents the condensate's momentum, and $\hat{J}_x$ bears direct information about the relative phase of the condensate's fractions in the left and right well.
The corresponding expectation values $\langle\cdot\rangle$  with respect to the quantum state are readily expressed with the help of the single-particle Bloch-vector 
\begin{equation}
\vec s=(s_x,s_y,s_z)=\frac{2}{N}\; (\langle \hat J_x\rangle,\langle \hat J_y\rangle,\langle \hat J_z\rangle).
\end{equation} 

In the mean-field or {\it classical} limit of large particle numbers $N$ (at fixed value $UN$) the dynamics of the condensate is described by the discrete Gross-Pitaevskii (GP) equation.
The main assumption underlying the latter is that the quantum state remains, at all times, a SU(2)- or atomic coherent state, i.e., a state of minimal uncertainties, with respect to the components of angular momentum (see, e.g., \cite{MCWW97,VA00}).
The coherence of the evolving state is sensitively measured by the single-particle- (or one-body) purity 
\begin{equation}
p=\frac{1}{2}(1+s_x^2+s_y^2+s_z^2),
\end{equation}
which ranges from $1/2$ to $1$ and takes the maximal value of one only for coherent states.
Thus, a value of $p<1$ indicates the departure from the mean-field, or classical, behavior.
Further information on the state is contained in its Wigner function $\rho_W$, a quasi-probability distribution evaluated on the spin-$j$ Bloch sphere as a function of the polar and azimuthal angles $\theta$ and $\phi$, respectively (see, e.g., \cite{Chuc10} and references therein).
For coherent states, the $\rho_W$ take a Gaussian shape, in particular they are positive.
A coherent state can thus be described by its center alone (a $c$-number), the evolution of which is given by the GP equation.
In angular momentum representation, the latter can be expressed with the help of the Bloch vector \cite{VA00}: 
\begin{equation}
\label{eq.GPE}
\dot{s}_x=-u\,s_ys_z,\,\, \dot{s}_y=s_z+u\,s_xs_z,\,\,\dot{s}_z=-\,s_y,
\end{equation}
where the time in (\ref{eq.GPE}) is rescaled by $t\rightarrow tK$ and describes the mean-field dynamics on a spin-$j$ Bloch sphere which is governed by the control parameter $u=UN/K$ \cite{BES90,MCWW97,SFGS97}.
Throughout the paper, we focus on the so-called Rabi regime of weak inter-atomic interactions $u \le 1$, where the atomic collisions do not yet induce self-trapping \cite{AGFHCO05} but notably influence the quantum dynamics \cite{MCWW97,VA00,ZNGO10} of the condensate, as the latter oscillates between the two wells.
Quantum and mean-field dynamics of the un-monitored system are revised further down in Sections \ref{sec.results_u0} and \ref{sec.results_u1}.

\begin{figure}
\begin{center}
\includegraphics[width=0.9\columnwidth,keepaspectratio,]{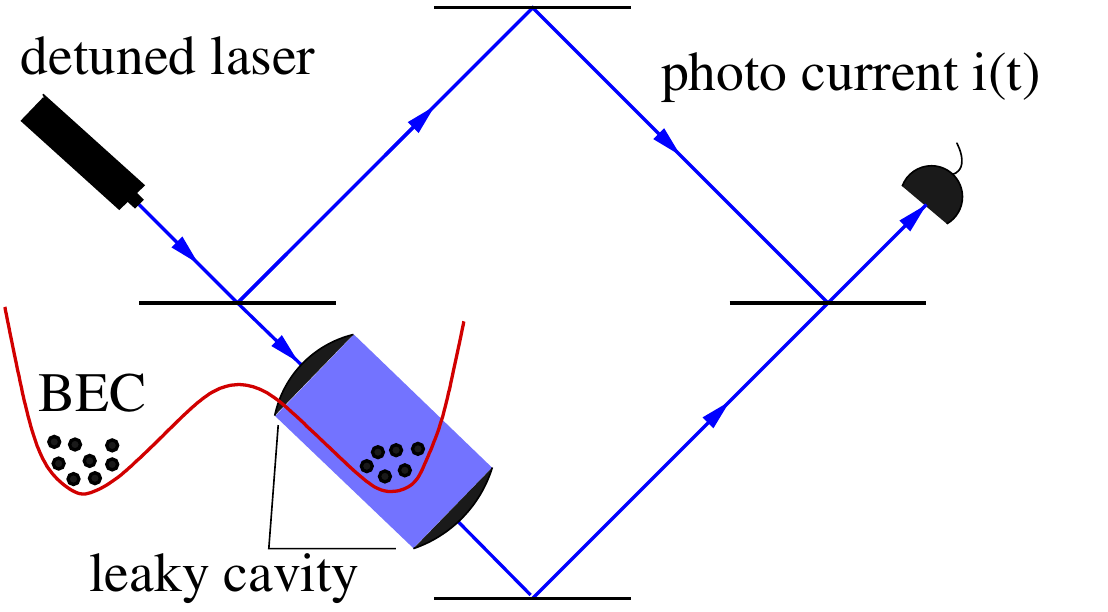}\hfill
\end{center}
\caption{(color online). 
Continuous unsharp detection of the atom number $n_2$ in the right well of a double-well potential:
the right well of the atom trap (red) is placed in a leaky cavity, which is illuminated by a continuous-wave laser detuned from atomic transitions.
Due to the atom-light scattering proportional to $n_2$, a relative phase shift is imprinted on the laser light which is subsequently detected by a Mach-Zehnder interferometer and read off as the photo current $i(t)$.
}
\label{setup}
\end{figure}

\section{Continuous measurements \label{sec.homodyne}}
Several proposals outlined the non-destructive measuring of a BEC via light- \cite{RW97,CorneyMilburn98,RW98,DDO02,MMR07,MR09,LL06,JJ11} or particle scattering \cite{SMH10,SOD11,Hunn12} including feedback control \cite{SHC10}.
Here, we focus on the scheme of Corney and Milburn \cite{CorneyMilburn98} sketched in Fig.~\ref{setup}.
In this setup, the number of atoms in {\it one}, say the right, site of a double-well potential is continuously and unsharply measured by placing the corresponding site inside a leaky optical cavity.
Well-detuned light from a continuous-wave laser pumps the cavity and is not absorbed by the atoms but suffers a phase shift proportional to the number { $n_2$ } of present atoms, which is detected by superposing the output light with a reference beam in a Mach-Zehnder interferometer (homodyning).
Also the atoms in the well experience a phase shift induced by the AC-Stark effect which is proportional to $\langle J_z\rangle$ and can thus be compensated by tilting the double-well potential in the earth gravitational field \cite{CorneyMilburn98}.

After adiabatic elimination of the degrees of freedom of the light field, the master equation for the reduced density operator $\hat\rho$ of the BEC reads \cite{Milburn.et.al94, CorneyMilburn98}:
\begin{equation}
\dot{\hat{\rho}} =  -i U[\hat{J}_z^2, \hat\rho] + i K [\hat{J}_x, \hat\rho] +
\frac{\gamma}{8} [\hat{J}_z,[\hat{J}_z, \hat\rho]]\,, 
\label{master}           
\end{equation}
with $\hbar\equiv1$. The strength of the measurement $\gamma = 64 \chi^2\epsilon^2/\Gamma^3$ is expressed in terms of the average
interaction strength  $\chi$ between atoms and light field in the cavity,  the amplitude $\epsilon$ of the coherent field
pumping the cavity and the cavity damping rate $\Gamma$.
For the measurement strength of $\gamma=0.01 K $ used below, the optical field strength amounts to $0.75 {\rm mW}$, where the mass of the bosons was assumed to be $1.5\times10^{-25}\rm{kg}$ \cite{CorneyMilburn98}.

The master equation (\ref{master}) describes the evolution of the mixed state associated with an {\it ensemble} of BECs which are prepared initially in the same state but then undergo evolutions with different measurement records, i.e., different detected photo currents (cp. setup in Fig.~\ref{setup}). 
The first two terms in (\ref{master}) correspond to the unitary evolution given by the Bose-Hubbard  Hamiltonian (\ref{bosehubb}) while the third term represents decoherence due to the interaction with the light beam.
This decoherence occurs with respect to the eigenbasis of the measured observable.
As the total number $N$ of atoms is fixed in this setup, a measurement of the number of atoms in one well is equivalent to a  measurement of $\hat{J}_z$.
Thus, the third term is the signature of the continuous measurement of the number of atoms in one well.

\subsection{Selective regime and stochastic Schr\"odinger equation}
In the following, we do not consider the mixed state that would result from averaging the possible outcomes of the continuous measurement. 
Instead, we are interested in the state evolution given a {\it particular} measurement record, i.e., a particular time-dependent photo current $i(t)$ induced in the photodetector (cp.~Fig.\ref{setup}).
In this so-called {\it selective} or {\it conditional} regime of measurement, i.e., conditioning the state evolution on a given photo current $i(t)$, an initially pure state $|\psi(t_0)\rangle $ of the condensate stays pure and can be described by a (stochastic) Schr\"odinger equation \cite{CorneyMilburn98}:
\begin{eqnarray}
d |\psi_c(t)\rangle &=&\left(-i \hat{H}_{BH}-\frac{\gamma}{8}\left(\hat{J}_z-\langle
\hat J_z\rangle_c \right)^2\right)dt\, |\psi_c(t)\rangle\nonumber\\
&{}&{} +\frac{{\gamma}}{2}(\hat{J}_z-\langle \hat J_z\rangle_c)(dI-\langle
\hat J_z\rangle_c dt)|\psi_c(t)\rangle\,,
\label{SSE}
\end{eqnarray}
where $\hat{H}_{BH}$ is the Bose-Hubbard  Hamiltonian (\ref{bosehubb}) and $\langle \hat J_z\rangle_c $ is the expectation value with respect to the
state  $|\psi_c(t)\rangle$.
The subscript $c$ marks expectation values and system states that are {\it conditioned} on a measurement record, i.e, conditioned on a specific photo current $i(t)$ measured until the time $t$.
Instead of $i(t)$ (cp.~Fig.\ref{setup}), we consider the (proportional) signal $I(t):=2 i(t)/ \gamma$, to keep our notation simple.
Then the increment of the measurement signal $I(t)$ from time $t$ to $t+dt$ reads:  
\begin{equation} 
dI=\langle
\hat J_z\rangle_c dt+\gamma^{-1/2}dW\,,
\label{moutput}
\end{equation}
where $dW$ is the increment of a white Gaussian noise process (so-called Wiener increment) \cite{Diosi88}, that reflects the measurement noise.
As in the master equation (\ref{master}), the first term in the stochastic evolution (\ref{SSE}) represents the Hamiltonian (unitary) dynamics while the second (nonlinear) term, corresponds to the double commutator in (\ref{master}).
As it is diagonal in $\hat J_z$, the latter term does not alter $\langle\hat{J}_z\rangle$, but reduces the expectation values of  $\hat{J}_x$ and $\hat{J}_y$.
The third, stochastic term --in combination with the second one--  narrows on average the variance of the measured observable $\hat{J}_z$.
To illustrate the action of the unsharp measurement with a simple example, we remark that in the {\it absence} of Hamiltonian dynamics, a continuous measurement projects the system (due to the second and third term of (\ref{SSE})) asymptotically into an eigenstate of the measured observable and this eigenstate can be identified by the measurement record \cite{Sondermann98}.

In the present case of a BEC evolving in a double-well potential, different energy scales compete.
From both, the master equation (\ref{master}) and the stochastic Schr\"odinger equation (\ref{SSE}), it is evident that terms corresponding to measurement and inter-atomic interactions increase quadratically with $N$ while the tunneling coupling depends linearly on $N$.
It is thus sensible to consider $\bar{\gamma}:=\gamma N/K$ as the normalized measurement strength with respect to the inter-well tunneling (accordingly, $U$ should be directly compared to $\gamma$).
The larger the value of $\bar\gamma$, the faster individual occupation levels $n_2$ can be resolved due to the measurement \cite{AudretschKonradScherer01,AudretschDiosiKonrad02}.
For $\bar\gamma  \gg1$, this results in the so-called shelving, i.e., while tunneling, the system remains longer on the individual levels $n_2$ and the tunneling oscillations are distorted
\footnote{The limit of $\protect\bar{\gamma} \rightarrow \infty$ corresponds to the Zeno regime, where the dynamics is brought to a halt.}.
In our present contribution, we concentrate on small to moderate measurement strengths $\bar\gamma$.
The exemplary value of $\bar\gamma=1$, used later, fulfills the operative definition that the tunneling frequency of the non-interacting system ($u=0$) be barely changed, i.e., that we are far away from the shelving regime (see Sec.\ref{sec.monitor_u0} further down).

As a side remark, let us note that similar physics, as described by the above master equation (\ref{master}), arises when
the noise in the quantum dynamics does not result from an unsharp measurement but from other sources like, e.g., interactions of the BEC with non-condensed atoms or by enforced stochastic driving \cite{KKV08,FSM10,BKOD09,KVC12}.

\section{State estimation\label{sec.estimate}}
According to Eq.~(\ref{moutput}), the measurement record $i(t)$ contains information about the expectation value of the measured observable $\hat J_z$, which can be extracted for sufficiently high measurement strength $\gamma$ using standard techniques such as Wiener filters \cite{Press.et.al99}, see, e.g., \cite{AudretschKonradScherer01}.   
Here, we would like to proceed beyond this point, namely, to infer (estimate) from a local observable $\hat J_z$ the {\it entire} quantum many-body state as it dynamically evolves under the influence of the measurement.
Moreover, this is to be accomplished in a single run, not in a state tomography experiment \cite{ST11,CR11} which requires repeated measurements on equally prepared systems.
We will assume that only the Bose-Hubbard Hamiltonian is known (i.e., the total particle number $N$ and the control parameter $u$) but not the initial state $|\psi(t_0)\rangle $ of the condensate.
We then continuously update an initial guess of the state $|\psi_e(t_0)\rangle $ according to the measurement results (i.e., the photo current $i(t)$) in order to obtain an estimate $|\psi_e(t)\rangle$ of the real state  $|\psi_c(t)\rangle$.  The overlap between both states is a measure for the fidelity of the estimate:
 \begin{equation}
F(t)=|\langle\psi_e(t)|\psi_c(t)\rangle|^2\,.
\label{eq.fidelity}
\end{equation}

For the special case of a continuously measured particle that maintains a Gaussian-shaped wave function, state estimation was discussed in \cite{DohertyJacobs99, Doherty.et.al00}.
The estimation scheme we consider here \cite{Diosi.et.al06} is versatile and can be employed in any continuous measurement scheme.
It is based on the It\^{o}-formalism and for ideal \cite{Diosi.et.al06} continuous measurements of otherwise closed, quantum systems, analytic arguments have been given that indicate the convergence of the estimate to the real state except for certain marginal cases \cite{Diosi.et.al06} (for a recent discussion on this convergence see, e.g., \cite{Rouchon11,AminiMazyarRouchon11}).
In this scheme, the state estimate is propagated in the same way as any real state conditioned on a given measurement record $I(t)$: 
\begin{eqnarray}
d |\psi_e(t)\rangle &=&\left(-i
\hat{H}_{BH}-\frac{\gamma}{8}\left(\hat{J}_z-\langle \hat J_z\rangle_e\right)^2\right)dt |\psi_e(t)\rangle\nonumber\\
&{}&{} +\frac{{\gamma}}{2}(\hat{J}_z-\langle \hat J_z\rangle_e)(dI-\langle \hat J_z\rangle_e dt) |\psi_e(t)\rangle\label{SSEE}.
\end{eqnarray}
At first sight, this equation appears to be identical to (\ref{SSE}) with the indices $c$ and $e$ exchanged.
Note, however, that the measurement signal $I(t)$ varies about the expectation value of $\hat{J}_z$ with respect to the {\it real state} rather
than to the state estimate.
This reflects the fact that $|\psi_e(t)\rangle$ is updated according to the measurement current $I(t)$ which is obtained with the system being in the real state.  
Therefore, the state estimate is {\it slaved} (i.e., linked) via the measurement signal $I(t)$ to the real state evolution.

\begin{figure}
\begin{center}
\includegraphics[width=0.99\columnwidth,keepaspectratio,type=pdf,ext=.pdf,read=.pdf]{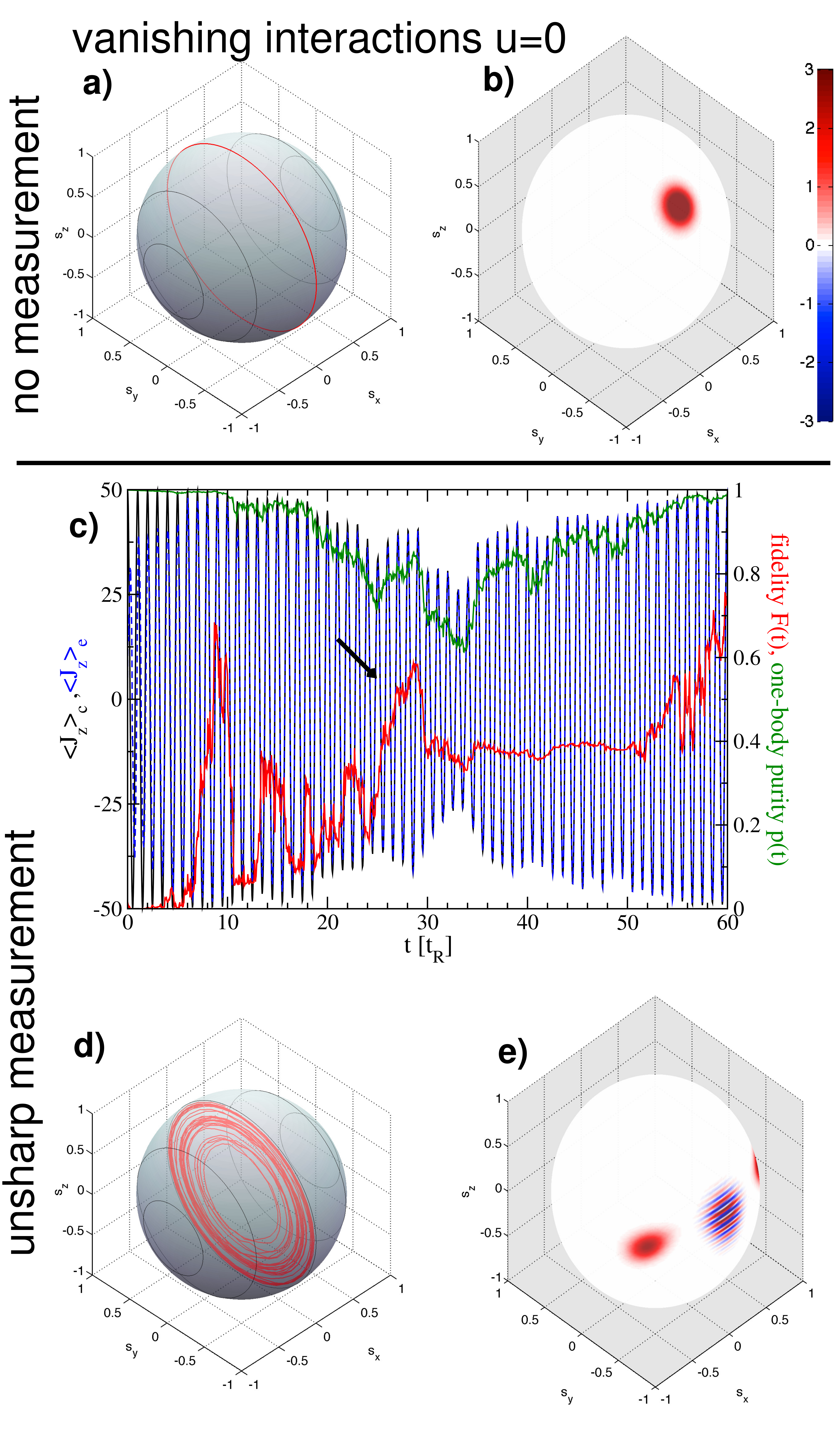}
\end{center}
\caption{(color online) System dynamics for vanishing interactions $u=0$
{\bf Upper panels:} a) The black lines indicate the mean-field phase space of a BEC in a double-well potential given by the discrete Gross-Pitaevskii equation (\ref{eq.GPE}) for vanishing interactions $u=0$ in the absence of measurement $\bar\gamma=0$.
The red line marks the evolution of the quantum mechanical Bloch vector $\vec s$ for $N=100$ particles, located initially in the left well (North pole of the Bloch sphere) and times $t=0-10$ Rabi periods $t_R$.
The Bloch vector remains (for all times) on the surface of the Bloch sphere, i.e., the state remains coherent throughout the evolution. 
b) The corresponding Wigner function $\rho_W$ after $\approx 4$ Rabi periods which remains Gaussian and positive (red areas) \cite{Note4}.
{\bf Middle panel:} The black (blue) line denotes the population imbalance $\langle \hat J_z\rangle_c$ ($\langle \hat J_z\rangle_e$) of the real (estimated) wave function. The red line indicates the fidelity $F(t)$ (\ref{eq.fidelity}) while the green line corresponds to the one-body purity $p(t)$.
All quantities are plotted in multiples of the Rabi period $t_R=2\pi/K$.
{\bf Lower panels.} d) Same as panel a) but now for non-vanishing measurement strength $\bar\gamma=1$ and times $t=10-40\, t_R$. 
The Bloch vector spirals towards the center of the sphere and is smallest at about $t=33t_R$ where also the one-body purity (see panel c) assumes its minimal value.
e) The corresponding Wigner function $\rho_W$ after $\approx 25$ Rabi periods (marked by a black arrow in panel c):
A strong non-coherent structure due to quantum interferences has emerged, in stark contrast to the case $\bar\gamma=0$ shown in panel b).}
\label{fig.results_u0}
\end{figure}

\begin{figure}
\begin{center}
\includegraphics[width=0.9\columnwidth,keepaspectratio,type=pdf,ext=.pdf,read=.pdf]{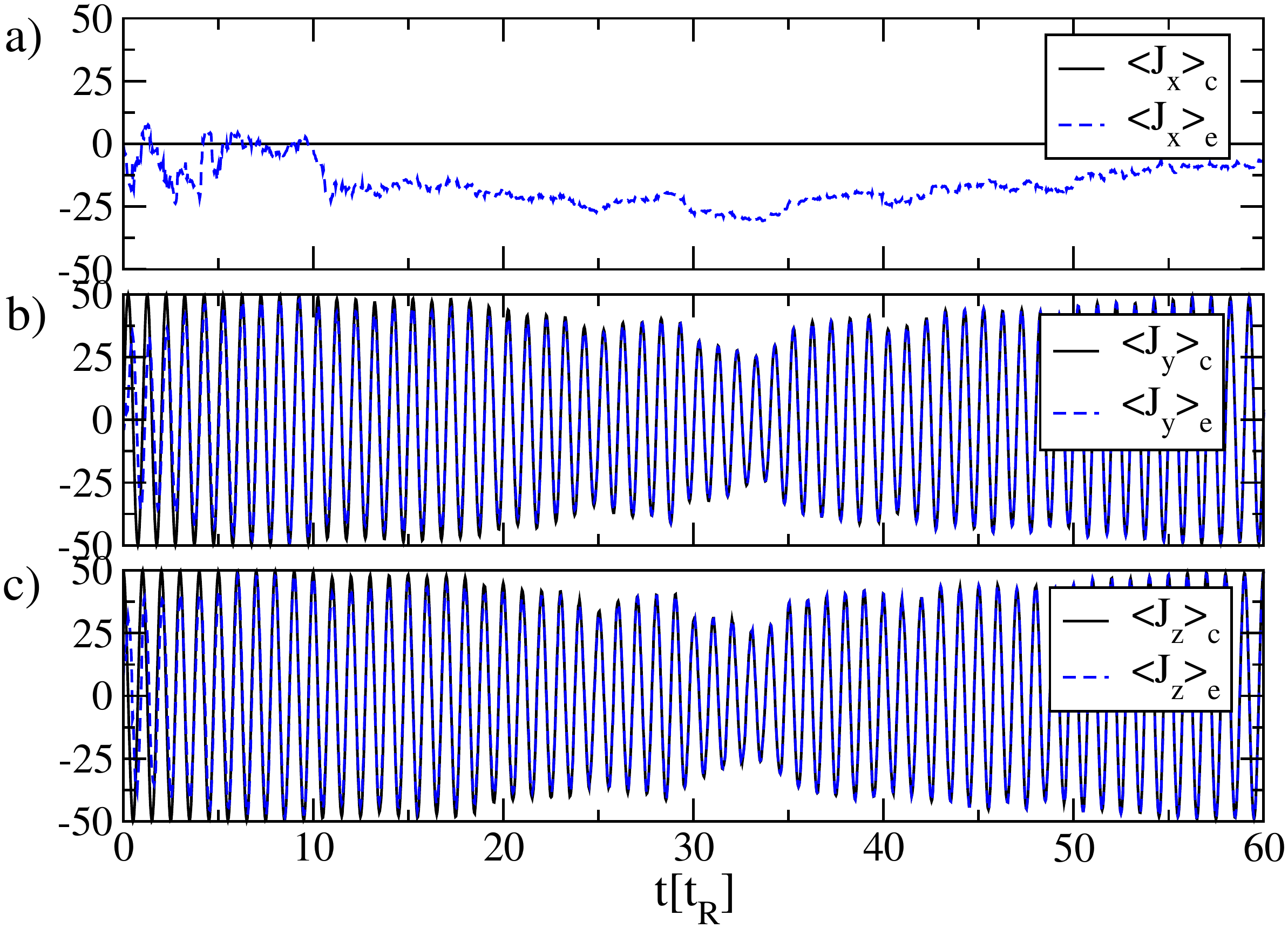}
\includegraphics[width=0.9\columnwidth,keepaspectratio,type=pdf,ext=.pdf,read=.pdf]{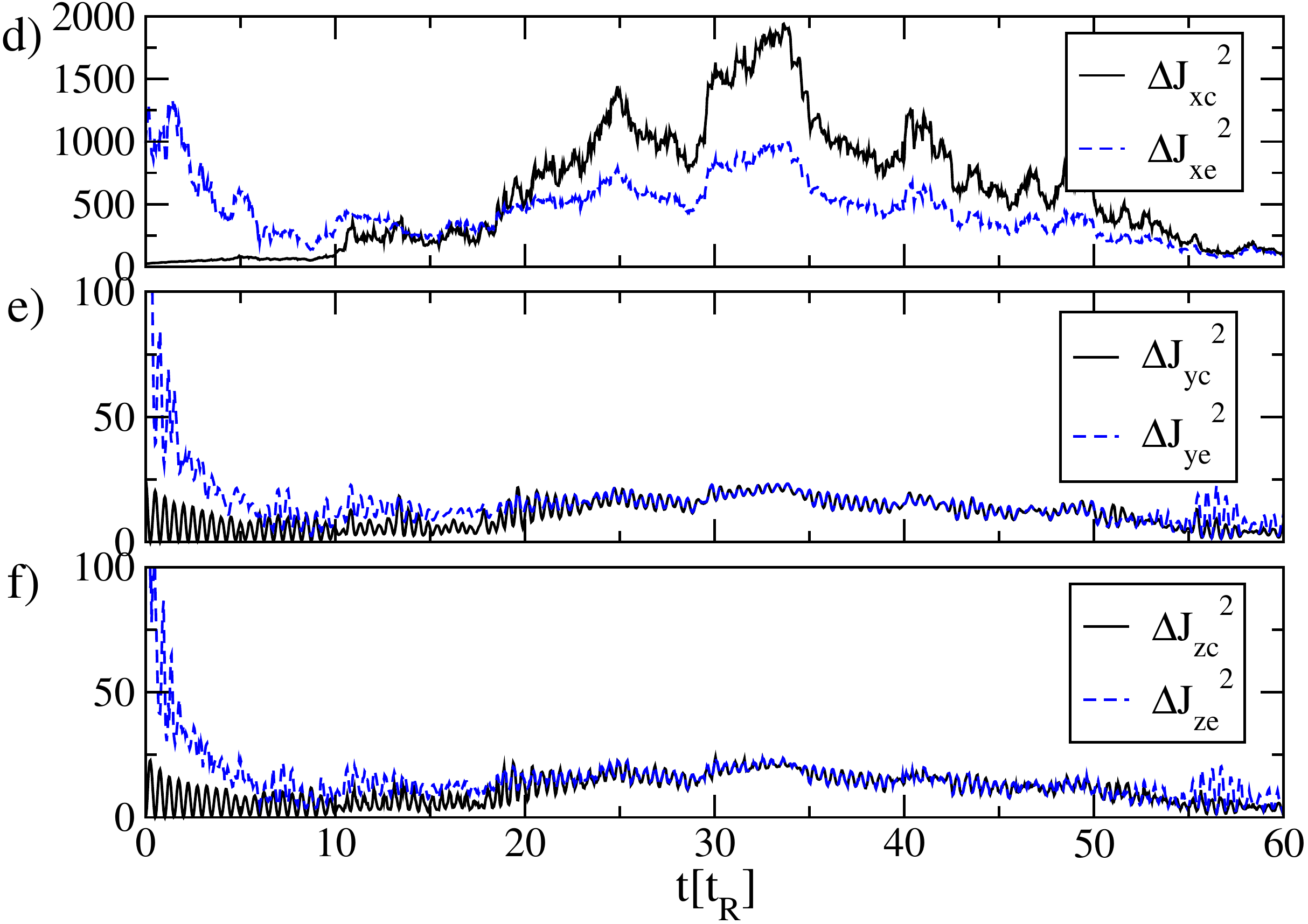}
\end{center}
\caption{(color online) Details of the system dynamics for vanishing interactions $u=0$
{\it Upper panels:}
As in Fig.~\ref{fig.results_u0}c), the black solid (blue dashed) lines denote the angular momentum expectation values $\langle \cdot\rangle_c$ ($\langle \cdot\rangle_e$) with respect to the real (estimated) wave function. The time is plotted in multiples of the Rabi period $t_R=2\pi/K$.
While $\langle \hat J_y\rangle_e$ and $\langle \hat J_z\rangle_e$ approach the values of the corresponding real wave function $|\psi_c(t)\rangle$, the estimate $\langle \hat J_x\rangle_e$ evolves stochastically.
{\it Lower panels:} The variances corresponding to the quantities in the upper panels are shown.
}
\label{fig.unsharp_u0_est}
\end{figure}

\begin{figure}
\begin{center}
\includegraphics[width=0.99\columnwidth,keepaspectratio,type=pdf,ext=.pdf,read=.pdf]{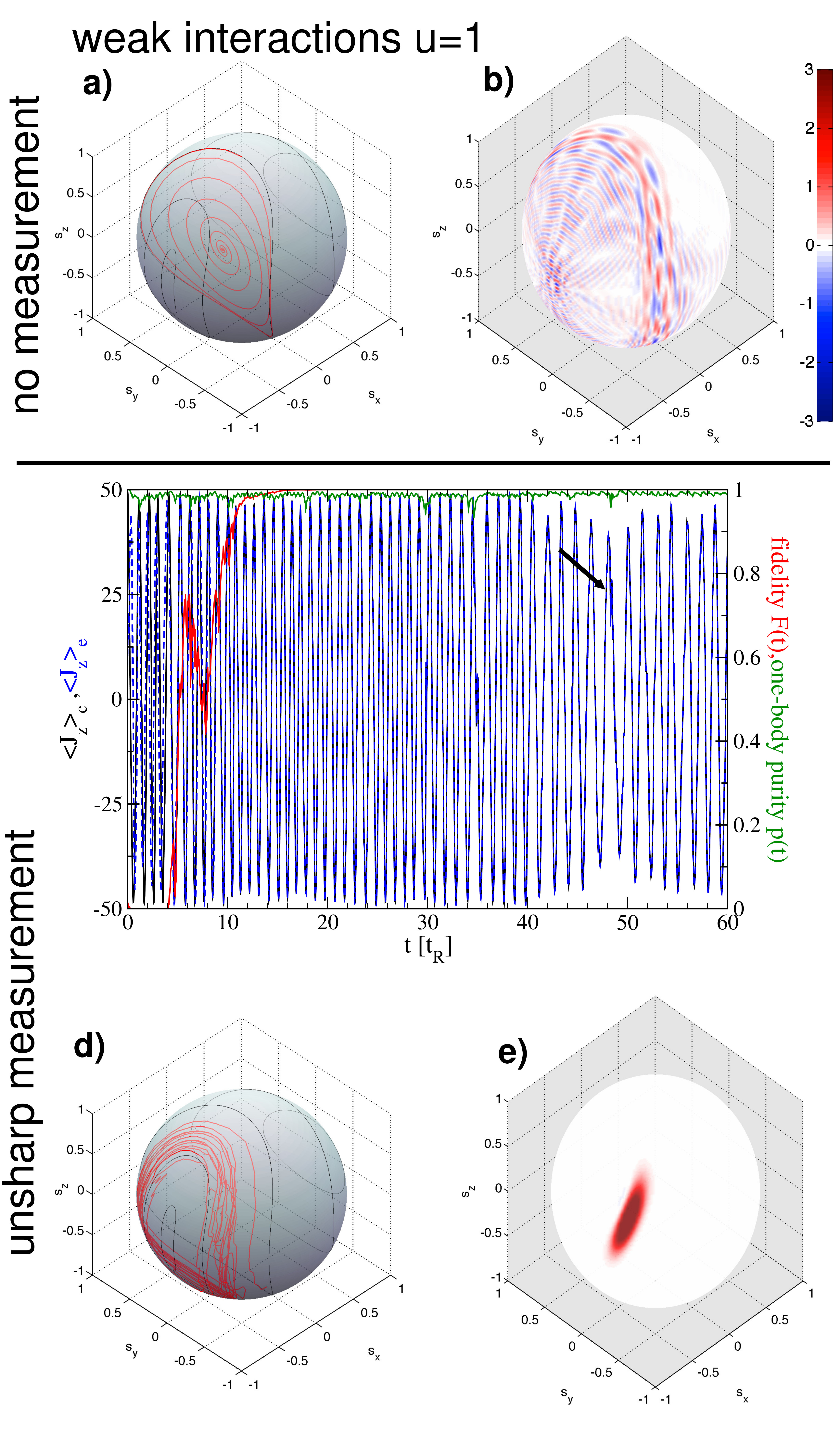}
\end{center}
\caption{(color online) System dynamics for weak interactions $u=1$
{\bf Upper panels:} a) The black lines indicate the mean-field phase space of a BEC in a double-well potential given by the discrete Gross-Pitaevskii equation (\ref{eq.GPE}) for weak interactions $u=1$ in the absence of measurement $\bar\gamma=0$.
The red line marks the evolution of the quantum mechanical Bloch vector $\vec s$ for $N=100$ particles, located initially in the left well (North pole of the Bloch sphere) and times $t=0-10$ Rabi periods.
The Bloch vector spirals into the interior of the Bloch sphere, indicating a loss of the one-body purity.
b) The corresponding Wigner function $\rho_W$ after $t=10 t_R$ develops an intricate structure with strong positive (red) and negative (blue) contributions \cite{Note4}. It is also smeared out along the central classical orbit of panel a) which results in $s_y=s_z=0$.
{\bf Middle panel:} The black (blue) line denotes the population imbalance $\langle \hat J_z\rangle_c$ ($\langle \hat J_z\rangle_e$) of the real (estimated) wave function. The red line indicates the fidelity $F(t)$ (\ref{eq.fidelity}) while the green line corresponds to the one-body purity $p(t)$.
{\bf Lower panels.} d) Same as panel a) but now for non-vanishing measurement strength $\bar\gamma=1$ and times $t=30-60\, t_R$. 
The Bloch vector essentially remains on the surface of the sphere and follows the mean-field (classical) evolution.
The corresponding one-body purity stays approximately at one, in stark contrast to the case $\bar\gamma=0$ shown in panel a).
e) The corresponding Wigner function $\rho_W$ after $\approx 48$ Rabi periods (marked by a black arrow in panel c): The latter remains essentially a coherent state, opposite to the case $\bar\gamma=0$ (panel b).
}
\label{fig.results_u1}
\end{figure}

\section{Results i: Vanishing inter-atomic interactions \label{sec.results_u0}}

We now have the tools at hand to study the effect of the unsharp measurement on the system dynamics and to explore whether we can successfully estimate the state of the entire system from the record of the local measurement on $\hat J_z$ alone.
At the focus of our work is how an initially coherent state is affected by the measurement, rather than the measurement-induced buildup of coherence studied, e.g., in \cite{RW97,CorneyMilburn98,DDO02}.
Hence, we concentrate our efforts on the exemplary case of all bosons being initially located in the first well, i.e., $|\psi_c(t=0)\rangle=|j\rangle$, an experimentally routinely prepared SU(2)-coherent state \cite{ZFG90}, located at the North pole of the Bloch sphere
\footnote{Together with the state $|-j\protect\rangle$ which represents a preparation on the South pole of the Bloch sphere, these are the only two Fock-states which are also coherent states, similar to the vacuum state of a harmonic oscillator.}.

As for the initial state of the {\it estimate} $|\psi_e(t=0)\rangle=\sum_{m=-j}^{m=j} c_m|m\rangle$, we pick the ``maximally uncertain" state, i.e., an equally weighted superposition of $0, 1, \ldots, N$ atoms in the first well $|c_m|^2=[2j+1]^{-1}$ with randomized relative phases $\arg(c_m)$.
Throughout all our numerical calculations, we fix the boson number $N=100$ and the inter-well tunneling coupling $K=1$.

\subsection{Un-monitored dynamics}
We start our discussion with the special and instructive case of vanishing inter-atomic interactions $u=0$ which was barely touched upon in \cite{CorneyMilburn98} and for which we first revise the un-monitored dynamics.
Without measurement $\bar\gamma=0$, the Bloch vector $\vec s$ of the system simply rotates at the Rabi frequency $K$ on the meridian defined by the intersection of the Bloch sphere with the $y-z$ plane \cite{VA00}, as shown in Fig.~\ref{fig.results_u0}a).
In this Rabi regime, the Wigner function $\rho_W$ associated with the quantum state, retains its Gaussian shape (cp. Fig.~\ref{fig.results_u0}b))
and its center moves on the corresponding mean-field trajectory, i.e., the respective solution of the GP equation (\ref{eq.GPE}) for $u=0$
\footnote{Animated movies of the (un-)monitored evolution for $u=0$ and $u=1$ can be found in the supplementary online material.}. 
Together with the lack of negative values, this points out  a ``classical state'' which possesses a joint probability distribution of the angles $\theta$ and $\phi$ and, hence, state estimation might be expected to work best for this case.

\subsection{Monitored dynamics\label{sec.monitor_u0}}%
The effect of the measurement on the quantum dynamics is shown in Figs.~\ref{fig.results_u0}c)-d) for the case $\bar\gamma=1$.
From panel c), we read off that the oscillation period of the population imbalance $\langle \hat J_z\rangle_c$ remains approximately the Rabi-period $t_R=2\pi/K$ (in accordance with our operative definition of a moderate measurement strength, see Sec. \ref{sec.homodyne}).
It is also evident, however, that the amplitude of the oscillations is not constant (as it is for $\bar\gamma=0$) but fluctuates. 
The decrease in oscillation amplitude is accompanied by a reduction of the one-body purity $p(t)$ (cp. Fig.~\ref{fig.results_u0}c)) which indicates that the wave function looses its Gaussian shape, i.e., departs from the mean-field behavior.
This is further corroborated by the Wigner function of the real state $|\psi_c(t)\rangle$ taken at about $t\approx25 t_R$ (cp. Fig.~\ref{fig.results_u0}e))
in which a prominent structure of both positive and negative contributions is evident. 
Furthermore, while $\rho_W$ stays well localized with respect to the polar angle (i.e., a small variance in $\theta$), there is no permanent localization in the perpendicular, i.e., in the azimuthal $\phi$-direction.
This spread over a part of the corresponding great circle on the Bloch sphere leads to a reduction of the length of the Bloch vector $\vec s_c$ and, thus, degrades $p(t)$ and the amplitude of the oscillations of $\langle\hat{J_z}\rangle_c$.
Summarizing these observations, one may thus say, that for vanishing inter-atomic interactions $u=0$, the measurement steers the system away from the mean-field behavior revised in the previous paragraph.

We present two complementary perspectives on this phenomenon, which we also observed for other ratios $\bar{\gamma}\le1$ of measurement and inter-well tunneling strengths:
in the Schr\"odinger picture, the unsharp measurement of the population imbalance reduces the variance of the corresponding observable $\hat J_z$ (cp. Fig.~\ref{fig.unsharp_u0_est}f)) due to the gain of information on its expectation value.
As a result of the unitary dynamics (which amounts to a bare rotation about the $x-$axis for $u=0$) this also yields a (systematic) reduction of the variance of $\hat J_y$ (cp. Fig.~\ref{fig.unsharp_u0_est}e)) but not of $\hat J_x$, i.e, it does not give rise to localization in the $x-$direction (cp. Fig.~\ref{fig.unsharp_u0_est}d).
Seen in the Heisenberg picture, the unsharp measurement reduces the variance of the time-dependent observable $\hat J_z(t)$ measured over a time period $\Delta t$.
In the case of vanishing atomic collisions $u=0$, the orbit of $\hat J_z(t)$ is given by $\{\sin(Kt) \hat{J}_y + \cos(Kt) \hat{J}_z|t\in \Delta t\}$, due to the rotation about the $x-$axis.
It does not include the angular momentum component $\hat{J}_x$ (the latter is constant and, in our case, its expectation value equals zero during the time evolution).
Thus, the variance in $\hat J_x$ is not (systematically) reduced but evolves in a predominantly stochastic fashion.
In the above example, the variance in $\hat J_x$ increases up to $t\approx33 t_R$, where $p(t)$ assumes its minimum, and then decreases again (cp.  solid line in Fig.~\ref{fig.unsharp_u0_est}d)).

As a somewhat crude mechanical analogy of this picture imagine a potter's wheel that spins about the $x-$axis and the chime representing the {\it uncertainty} of the quantum state in the three directions.
The effect of the unsharp measurement is like softly compressing the chime in $z-$direction: due to the wheel's rotation, one will finally obtain a piece of chime that is cylindrical, with small radius (i.e., uncertainty) in the $y-z$-plane but increased height in the $x-$direction.
One aspect which is not captured by this mechanical picture is that, due to the stochastic character of the unsharp measurement, there is an {\it additional} random component in the evolution of the chime's shape.
That is, the uncertainty in $\hat{J}_x$ is not monotonically growing and the one in $\hat{J}_y$ and $\hat{J}_z$ is not monotonically decreasing.

\subsection{State estimation}
We have yet to discuss the performance of the estimator (\ref{SSEE}) to infer the many-body state of the system from the measured photo current $i(t)$. 
From the exemplary case shown in Fig.~\ref{fig.results_u0}, we find that the oscillations in $\langle \hat J_z\rangle_c$ themselves are, after a short transition time, well reproduced by the estimator $\langle \hat J_z\rangle_e$.
Nevertheless, the estimated state $|\psi_e(t)\rangle$ does not fully converge to the real state $|\psi_c(t)\rangle$ within the considered time period of $60$ Rabi periods, as spelled out by the fidelity that stays below unity (cp Fig.~\ref{fig.results_u0}c)).
Given the fact that the un-monitored BEC dynamics is ``simple" Rabi oscillations in the population imbalance, this is somewhat surprising.
Moreover, the fidelity oscillates quite strongly, {\it despite} the rather good agreement of the real $\langle \hat J_z\rangle_c$ and the estimated  $\langle \hat J_z\rangle_e$ particle imbalance.
An inspection of the estimated state reveals (cp. Figs.~\ref{fig.unsharp_u0_est}a)-c)) that indeed $\langle \hat J_y\rangle_e$ and $\langle \hat J_z\rangle_e$ agree quite well with the corresponding components of the real state while $\langle \hat J_x\rangle_e$ behaves erratically.
We can immediately understand this behavior by means of the above discussed evolution in the Heisenberg picture: due to the rotation about the $x-$axis, the information on $\hat J_x$ is not available in the measurement and, hence, the estimated $\langle \hat J_x\rangle_e$ does not converge to the real result $\langle \hat J_x\rangle_c$.

\section{Results ii: Weak inter-atomic interactions\label{sec.results_u1}}

\subsection{Un-monitored dynamics}
We now turn to the more general scenario of non-vanishing but weak inter-atomic interactions $u\le1$.
In this case, the mean-field (i.e., the classical) approximation deviates substantially from the quantum dynamics, already in the {\it absence} of measurements (see, e.g. \cite{MCWW97,VA00}):
the mean-field solutions are again closed trajectories on the Bloch sphere with pronounced oscillations in the particle imbalance $s_z$ (see black lines in Fig.~\ref{fig.results_u1}a)).
Due to the inter-atomic interactions, the dynamics is not merely a tunneling-induced Rabi-like rotation about a fixed (here the $x$-) axis but there is an additional  ``nonlinear rotation''  about the $z$ axis with a rotation angle depending on the $z$-component (see Hamiltonian (\ref{bosehubb})) and a different rotation frequency for each trajectory
\footnote{The nonlinear rotation is manifest in the ``curved'' mean-field trajectories for $u=1$ in contrast to the ``straight'' trajectories that result from  plane-sphere intersections for $u=0$ (cp. thin black lines in Fig.~\ref{fig.results_u0}a) and Fig.~\ref{fig.results_u1}a))}.
Since the quantum state for a finite number $N$ of atoms is not a single point on the Bloch sphere, but has a width according to its Wigner function, the different evolution frequencies lead to a dephasing in the course of the tunneling dynamics.
As a result, the oscillations in the atomic population imbalance are damped and cease after a few cycles, even in the regime of weak inter-atomic interactions $u=1$.
In this process, the quantum mechanical Bloch vector $\vec s$ spirals onto the $x-$axis of the Bloch sphere which implies that eventually $\langle \hat J_y\rangle=\langle \hat J_z\rangle=0$ (see red line in Fig.~\ref{fig.results_u1}a))
\footnote{A detailed discussion of the asymptotic behavior of the Bloch vector for various initial conditions and interaction values can be found in \cite{Chuc10}.}.
The corresponding Wigner function (cp.\ Fig.~\ref{fig.results_u1}b)) assumes positive (red) and negative (blue) values and, after several Rabi periods $t_R$, is almost symmetrically distributed about the $x$-axis which again implies $\langle \hat J_z\rangle=0$, and thus the collapse of the oscillations.

\subsection{Monitored dynamics}
It may come as a surprise that the impact of the unsharp measurement of $\hat{J}_z$ is to restore the oscillations, even for moderate strengths $\bar{\gamma}=1$ (cp.\ Fig.~\ref{fig.results_u1}c)) 
\footnote{We found that $\protect\bar{\gamma}$ should be greater or equal to $u$ to prevent the damping of the Rabi oscillations but of the order of $K$ in order to not significantly disturb them.
A quantitative analysis of the minimal measurement strength should be based on a comparison of the dephasing rate (specific to the state under investigation at a given value of $u$) and the measurement strength $\protect\bar{\gamma}$.
}.
This effect was first mentioned in \cite{CorneyMilburn98,RW98} and we would like to add to its understanding with a detailed analysis supplemented by a mean-field perspective before we proceed to the entirely new results on the state estimation.

We start with the observation that in the presence of inter-atomic interactions $u\neq0$, the unsharp measurement {\it induces} classical behavior of the BEC:
(i) in contrast to the un-monitored quantum dynamics for $u=1$ discussed in the beginning of this section, under the unsharp measurement (e.g.,  $\bar\gamma=u=1$), the Bloch vector $\vec s_c$ essentially stays at the surface of the Bloch sphere (cp. Fig.~\ref{fig.results_u1}d)) with $p(t)\approx1$  (cp. green line in Fig.~\ref{fig.results_u1}c)).
That is, the BEC stays approximately in a coherent state of minimal uncertainty with respect to the angular momentum components.
This is further corroborated by our observation, that the Wigner function of the BEC is positive and approximately Gaussian (cp. Fig.~\ref{fig.results_u1}e)).
To draw the connection to the above discussed mechanical analog, the spinning axis of the potter's wheel is now changing in time, due to the nonlinear rotation caused by the nonlinear term in (\ref{bosehubb}).
As a result, the variances of {\it all} three angular momentum components are systematically reduced (still subject, however, to fluctuations that result from the stochastic character of the measurement).

This has to be contrasted with the case of vanishing interactions $u=0$: here the un-monitored dynamics follows the classical evolution with $p(t)=1$ but 
once the measurement is turned on, becomes truly quantum as indicated by a non-positive Wigner function and a fluctuating one-body purity $p(t)$.

(ii) Secondly, we find that this approximately coherent state evolves in the vicinity of the classical (mean-field) trajectories (cp. Fig.~\ref{fig.results_u1}d)).
In other words, by means of continuous unsharp measurements, classical trajectories can be realized, that would rapidly dephase for $\bar\gamma=0$.
In this sense, the unsharp measurement can be considered weak --as it does not considerably change the mean-field dynamics. 
But, at the same time, it has a pronounced impact on the quantum dynamics, namely, to halt the dephasing.

Let us, however, point out that, as a result of the stochasticity introduced by the measurement, the quantum state does not evolve on a single mean-field trajectory, i.e., a line on the Bloch sphere, but explores an entire area (cp. Fig.~\ref{fig.results_u1}d)).
As a side remark, we note that this can as well be inferred from the time evolution of $\langle \hat J_z\rangle_c$ shown in Fig.~\ref{fig.results_u1}c):
as mentioned above, in the presence of interatomic interactions $u\neq0$, different mean-field trajectories come with different frequencies \cite{Chuc10}.
We should thus observe that the oscillation in the particle imbalance $\langle \hat J_z\rangle_c$ are restored \cite{CorneyMilburn98,RW98} but not necessarily at the same period. 
Indeed, we find that the frequency of the oscillations in Fig.~\ref{fig.results_u1}c) changes during the course of time (see, e.g., around $t=50t_R$), while this is not the case for $u=0$ (cp. Fig.~\ref{fig.results_u0}c)).
We stress, that the smaller oscillation amplitude around $t=50 t_R$ corresponds to an orbit located at the $s_x<0$ hemisphere, as can be seen from Fig.~\ref{fig.results_u1}d).
Specifically, it is not due to a dephasing as for $u=0$ since, in contrast to Fig.~\ref{fig.results_u0}c), the one-body-purity stays close to one.

\subsection{State estimation}

After discussing the underlying mechanism responsible for halting the dephasing, we now turn to the performance of the estimator (\ref{SSEE}).
In contrast to the case $u=0$, not only are the oscillations of $\langle \hat J_z\rangle_c$ well described by estimate $\langle \hat J_z\rangle_e$ but also the fidelity increases within a few cycles to unity (cp. Fig.~\ref{fig.results_u1}c)), indicating the convergence of the estimated state $|\psi_e(t)\rangle$ to the real state $|\psi_c(t)\rangle$.
That is, a local measurement of the particle imbalance $\hat J_z$ leads to information on the entire wave function $|\psi_c(t)\rangle$.
To understand this, we return to the Heisenberg picture, discussed above in Section \ref{sec.results_u0}:
as a result of the nonlinear rotation for $u\neq 0$, the set of measured observables is no longer restricted to the $y-z$-plane but also contains $\hat{J}_x$.
Hence, the measurement of $\hat J_z$ yields direct information on all angular momentum components and, thus, we obtain  
an {\it informationally complete} \cite{BuschGrabowskiLahti95} set of observables.
That is, sufficient information can be gathered from the continuous unsharp measurement to determine the state of the system.
Indeed, without atomic collisions ($u=0$) where the orbit of $\hat{J}_z(t)$ does not contain $\hat{J}_x$, the estimate does not converge towards the real state within the considered time period.

We emphasize, that the complete knowledge of the wave function was gained via the measurement of a {\it single} observable on a single realization of the system consisting of hundred atoms with a Hilbert space of dimension $d=101$. 
This is remarkable because a full state tomography (by means of von Neumann projection measurements) of such a system would require measurements of $d^2-1= 10200$ linear independent observables, each one carried out on a large ensemble of equally prepared systems
\footnote{See \cite{ST11,CR11} and references therein, for recent advances in the tomography of a BEC in a double-well potential}.

\section{Summary and discussion}\label{discussion}

In summary, we studied the continuous unsharp measurement of the atom number in {\it one} site of a double-well potential loaded with a BEC.
Based on the setup proposed in \cite{CorneyMilburn98} we focused on the selective regime of measurement.
In a first step, we investigated the interplay between internal many-body dynamics and the measurement for the case of an initially coherent state of maximal particle imbalance.
We showed that in the absence of inter-atomic interactions, an unsharp measurement of moderate strength steers the system away from the coherent mean-field evolution of the un-monitored BEC, and leads to a genuine quantum state manifest in a Wigner function of oscillating sign.
Contrary, for weak inter-atomic interactions ($u=1$) the un-monitored dynamics rapidly deviates from the mean-field behavior \cite{MCWW97,VA00} and the inter-well tunneling oscillations of the BEC are damped out.
As first noted in \cite{CorneyMilburn98,RW98}, the latter can be restored by a measurement of moderate strength $\bar{\gamma}$.
We explained the underlying mechanism and showed that the monitored dynamics evolves essentially as a coherent state, i.e., close to the surface of the Bloch sphere and in the vicinity of the solutions that correspond to the Gross-Pitaevskii equation.
Hence, we demonstrated that the unsharp measurement can induce quantum or classical (mean-field) behavior depending on the strength of the inter-atomic interactions. 

Secondly, we went beyond the standard measurement scheme and explored the viability to infer from the measurement record not only the expectation value of the unsharply measured observable but the entire system state \cite{Diosi.et.al06} in a {\it single} run of the experiment and {\it without} any knowledge of the initial state.
Opposite to the naive expectation, this was not achieved for $u=0$ but in the presence of weak inter-atomic interactions $u\neq0$ and sufficiently high but still moderate measurement strength $\bar\gamma$.
Specifically, we demonstrated for the exemplary case of $u=1$ that, after a certain waiting period, the time-evolving state of the atoms can be monitored with perfect fidelity.
This convergence in the course of a continuous measurement was explained in terms of an informationally complete set of observables that arises from the interactions.

The observations for $u\neq0$, show  that the continuous measurement of moderate strength $\bar\gamma=1$, modifies the dynamics considerably as it compensates the dephasing and induces classical (mean-field) dynamics.
On the other hand, it does not alter the structure of the mean-field dynamics itself.
That is, combined with our estimation procedure, the unsharp measurement can be used to prepare a non-dispersing coherent wave packet that evolves close to the mean-field solutions and, at the same time, monitor the state with perfect fidelity.
This should be contrasted with recent studies on periodically \cite{BKOD09} or stochastically \cite{KKV08,KVC12} driven BECs in double-well potentials, 
where due to a strong driving, the dephasing of an initially coherent state is reduced but at the same time, the entire dynamics is brought to a halt.
It is as well different from the dynamical localization observed for atoms in periodically driven lattices (see, e.g., \cite{Ecke_etal09}) where by appropriate choice of the driving parameter the inter-site tunneling is effectively switched off.
As a last example for the mitigation of dephasing, we mention non-dispersing wave packets that have been observed in the context of Rydberg atoms \cite{HL95, BDZ02,DMR09} and which may actually \emph{exhibit} dynamics.
In that case, however, the dephasing is suppressed due to an external driving which typically alters the underlying classical phase space considerably.

Let us discuss the convergence of the estimate and the emergence of classical properties from the perspective of measurement theory.
The arguments given in \cite{Diosi.et.al06}, supported by our numerical evidence for $u\neq0$, indicate that the convergence of the estimated state to the real state --conditioned on a particular measurement record $I(t)$-- occurs independently of the choice of the initial estimate $|\psi_e(t=0)\rangle$.
If so, also all real initial states conditioned on the same $I(t)$ converge to the same state $|\psi_c(t)\rangle$  given by the measurement record since estimated and conditioned states propagate according to the same evolution equation, cp.\ Eqs.\ (\ref{SSE}) and (\ref{SSEE}).    
Rather than state monitoring, this projection of the set of initial states onto a single state resembles a dynamical state preparation procedure.
The latter enforces a reduction of complexity in the sense that (after the projection has occurred) there is a one-to-one correspondence between the  classical information contained in the measurement record $I(t)$ and the in principle complex quantum information contained in the state $|\psi_c(t)\rangle$. 
In the monitored dynamics, this reduction of complexity is manifest in the observed transition from quantum to classical properties, i.e., the BEC evolves approximately in a coherent state.

Experimentally, BECs in a double-well potential have been thoroughly studied \cite{AGFHCO05,MO06,BKOD09,ZNGO10,Gross_etal10} and the effective inter-atomic interaction $u$ is under precise control.
The investigated unsharp measurement thus represents a prototype experiment which would allow to observe the transition from an informationally incomplete measurement (at  $u=0$) to the informationally complete case (at weak inter-atomic interactions) by tuning the internal system dynamics.
Given the estimated laser intensity of the probe field to be $0.75 {\rm mW}$ for $N=100$ bosons, these experiments should be realizable with state-of-the-art technologies.

As a future direction of this research line, one may explore the possibility to estimate the {\it total} number of atoms in a BEC placed in a double well by measuring only the number of atoms in one well.
Furthermore, the general concept of quantum to classical transitions, as induced by the measurement, should be tested for different systems.
Finally, the full knowledge of the BEC's state in the presence of dynamics may be employed to control the system by unitary feedback depending on the measurement record.

\begin{acknowledgments}
We acknowledge financial support by the South Africa/Germany Research Cooperation Programme NRF Grant 69436 and the BMBF, Grant 08/008. In addition T.K. is grateful for support by the NRF Focus Area Grant 65579. M.H. and A.B. acknowledge support by the EU COST Action MP1006 'Fundamental
Problems in Quantum Physics' and DFG Research Unit 760.
\end{acknowledgments}

%

\bibliography{measurement}

\begin{thebibliography}{70}%
\makeatletter
\providecommand \@ifxundefined [1]{%
 \@ifx{#1\undefined}
}%
\providecommand \@ifnum [1]{%
 \ifnum #1\expandafter \@firstoftwo
 \else \expandafter \@secondoftwo
 \fi
}%
\providecommand \@ifx [1]{%
 \ifx #1\expandafter \@firstoftwo
 \else \expandafter \@secondoftwo
 \fi
}%
\providecommand \natexlab [1]{#1}%
\providecommand \enquote  [1]{``#1''}%
\providecommand \bibnamefont  [1]{#1}%
\providecommand \bibfnamefont [1]{#1}%
\providecommand \citenamefont [1]{#1}%
\providecommand \href@noop [0]{\@secondoftwo}%
\providecommand \href [0]{\begingroup \@sanitize@url \@href}%
\providecommand \@href[1]{\@@startlink{#1}\@@href}%
\providecommand \@@href[1]{\endgroup#1\@@endlink}%
\providecommand \@sanitize@url [0]{\catcode `\\12\catcode `\$12\catcode
  `\&12\catcode `\#12\catcode `\^12\catcode `\_12\catcode `\%12\relax}%
\providecommand \@@startlink[1]{}%
\providecommand \@@endlink[0]{}%
\providecommand \url  [0]{\begingroup\@sanitize@url \@url }%
\providecommand \@url [1]{\endgroup\@href {#1}{\urlprefix }}%
\providecommand \urlprefix  [0]{URL }%
\providecommand \Eprint [0]{\href }%
\providecommand \doibase [0]{http://dx.doi.org/}%
\providecommand \selectlanguage [0]{\@gobble}%
\providecommand \bibinfo  [0]{\@secondoftwo}%
\providecommand \bibfield  [0]{\@secondoftwo}%
\providecommand \translation [1]{[#1]}%
\providecommand \BibitemOpen [0]{}%
\providecommand \bibitemStop [0]{}%
\providecommand \bibitemNoStop [0]{.\EOS\space}%
\providecommand \EOS [0]{\spacefactor3000\relax}%
\providecommand \BibitemShut  [1]{\csname bibitem#1\endcsname}%
\let\auto@bib@innerbib\@empty
\bibitem [{\citenamefont {Plenio}\ and\ \citenamefont
  {Knight}(1998)}]{PlenioKnight98}%
  \BibitemOpen
  \bibfield  {author} {\bibinfo {author} {\bibfnamefont {M.}~\bibnamefont
  {Plenio}}\ and\ \bibinfo {author} {\bibfnamefont {P.}~\bibnamefont
  {Knight}},\ }\href@noop {} {\bibfield  {journal} {\bibinfo  {journal} {Rev.
  Mod. Phys.}\ }\textbf {\bibinfo {volume} {70}},\ \bibinfo {pages} {101}
  (\bibinfo {year} {1998})}\BibitemShut {NoStop}%
\bibitem [{\citenamefont {Jacobs}\ and\ \citenamefont
  {Steck}(2006)}]{JacobsSteck06}%
  \BibitemOpen
  \bibfield  {author} {\bibinfo {author} {\bibfnamefont {K.}~\bibnamefont
  {Jacobs}}\ and\ \bibinfo {author} {\bibfnamefont {A.}~\bibnamefont {Steck}},\
  }\href@noop {} {\bibfield  {journal} {\bibinfo  {journal} {Contemp. Phys.}\
  }\textbf {\bibinfo {volume} {47}},\ \bibinfo {pages} {279} (\bibinfo {year}
  {2006})}\BibitemShut {NoStop}%
\bibitem [{\citenamefont {Wiseman}\ and\ \citenamefont
  {Milburn}(2009)}]{WisemanMilburn09}%
  \BibitemOpen
  \bibfield  {author} {\bibinfo {author} {\bibfnamefont {H.}~\bibnamefont
  {Wiseman}}\ and\ \bibinfo {author} {\bibfnamefont {G.~J.}\ \bibnamefont
  {Milburn}},\ }\href@noop {} {\emph {\bibinfo {title} {Quantum Measurement and
  Control}}}\ (\bibinfo  {publisher} {Cambridge University Press},\ \bibinfo
  {year} {2009})\BibitemShut {NoStop}%
\bibitem [{\citenamefont {Barchielli}\ and\ \citenamefont
  {Gregoratti}(2009)}]{Barchielli09}%
  \BibitemOpen
  \bibfield  {author} {\bibinfo {author} {\bibfnamefont {A.}~\bibnamefont
  {Barchielli}}\ and\ \bibinfo {author} {\bibfnamefont {M.}~\bibnamefont
  {Gregoratti}},\ }\href@noop {} {\emph {\bibinfo {title} {Quantum \newline
  Trajectories and Measurements in Continuous Time}}}\ (\bibinfo  {publisher}
  {Springer},\ \bibinfo {year} {2009})\BibitemShut {NoStop}%
\bibitem [{\citenamefont {Caves}\ and\ \citenamefont
  {Milburn}(1987)}]{CavesMilburn87}%
  \BibitemOpen
  \bibfield  {author} {\bibinfo {author} {\bibfnamefont {C.}~\bibnamefont
  {Caves}}\ and\ \bibinfo {author} {\bibfnamefont {G.}~\bibnamefont
  {Milburn}},\ }\href@noop {} {\bibfield  {journal} {\bibinfo  {journal}
  {Phys.\ Rev.}\ }\textbf {\bibinfo {volume} {A 36}},\ \bibinfo {pages} {5543}
  (\bibinfo {year} {1987})}\BibitemShut {NoStop}%
\bibitem [{\citenamefont {Audretsch}\ \emph
  {et~al.}(2002{\natexlab{a}})\citenamefont {Audretsch}, \citenamefont
  {Konrad},\ and\ \citenamefont {Scherer}}]{AudretschKonradScherer02}%
  \BibitemOpen
  \bibfield  {author} {\bibinfo {author} {\bibfnamefont {J.}~\bibnamefont
  {Audretsch}}, \bibinfo {author} {\bibfnamefont {T.}~\bibnamefont {Konrad}}, \
  and\ \bibinfo {author} {\bibfnamefont {A.}~\bibnamefont {Scherer}},\
  }\href@noop {} {\bibfield  {journal} {\bibinfo  {journal} {Phys. Rev. A}\
  }\textbf {\bibinfo {volume} {65}},\ \bibinfo {pages} {033814} (\bibinfo
  {year} {2002}{\natexlab{a}})}\BibitemShut {NoStop}%
\bibitem [{\citenamefont {Korotkov}(2000)}]{Korotkov00a}%
  \BibitemOpen
  \bibfield  {author} {\bibinfo {author} {\bibfnamefont {A.}~\bibnamefont
  {Korotkov}},\ }\href@noop {} {\bibfield  {journal} {\bibinfo  {journal}
  {Physica B}\ }\textbf {\bibinfo {volume} {280}},\ \bibinfo {pages} {412}
  (\bibinfo {year} {2000})}\BibitemShut {NoStop}%
\bibitem [{\citenamefont {Audretsch}\ \emph {et~al.}(2001)\citenamefont
  {Audretsch}, \citenamefont {Konrad},\ and\ \citenamefont
  {Scherer}}]{AudretschKonradScherer01}%
  \BibitemOpen
  \bibfield  {author} {\bibinfo {author} {\bibfnamefont {J.}~\bibnamefont
  {Audretsch}}, \bibinfo {author} {\bibfnamefont {T.}~\bibnamefont {Konrad}}, \
  and\ \bibinfo {author} {\bibfnamefont {A.}~\bibnamefont {Scherer}},\
  }\href@noop {} {\bibfield  {journal} {\bibinfo  {journal} {Phys. Rev. A}\
  }\textbf {\bibinfo {volume} {63}},\ \bibinfo {pages} {052102} (\bibinfo
  {year} {2001})}\BibitemShut {NoStop}%
\bibitem [{\citenamefont {Audretsch}\ \emph {et~al.}(2007)\citenamefont
  {Audretsch}, \citenamefont {Klee},\ and\ \citenamefont
  {Konrad}}]{AudretschKleeKonrad04}%
  \BibitemOpen
  \bibfield  {author} {\bibinfo {author} {\bibfnamefont {J.}~\bibnamefont
  {Audretsch}}, \bibinfo {author} {\bibfnamefont {F.}~\bibnamefont {Klee}}, \
  and\ \bibinfo {author} {\bibfnamefont {T.}~\bibnamefont {Konrad}},\
  }\href@noop {} {\bibfield  {journal} {\bibinfo  {journal} {Phys. Lett. A}\
  }\textbf {\bibinfo {volume} {361}},\ \bibinfo {pages} {212} (\bibinfo {year}
  {2007})}\BibitemShut {NoStop}%
\bibitem [{\citenamefont {Silberfarb}\ \emph {et~al.}(2005)\citenamefont
  {Silberfarb}, \citenamefont {Jessen},\ and\ \citenamefont
  {Deutsch}}]{Silberfarb.et.al.05}%
  \BibitemOpen
  \bibfield  {author} {\bibinfo {author} {\bibfnamefont {A.}~\bibnamefont
  {Silberfarb}}, \bibinfo {author} {\bibfnamefont {P.~S.}\ \bibnamefont
  {Jessen}}, \ and\ \bibinfo {author} {\bibfnamefont {I.~H.}\ \bibnamefont
  {Deutsch}},\ }\href@noop {} {\bibfield  {journal} {\bibinfo  {journal} {Phys.
  Rev. Lett.}\ }\textbf {\bibinfo {volume} {95}},\ \bibinfo {pages} {030402}
  (\bibinfo {year} {2005})}\BibitemShut {NoStop}%
\bibitem [{\citenamefont {Smith}\ \emph {et~al.}(2006)\citenamefont {Smith},
  \citenamefont {Silberfarb}, \citenamefont {Deutsch},\ and\ \citenamefont
  {Jessen}}]{Smith.et.al.06}%
  \BibitemOpen
  \bibfield  {author} {\bibinfo {author} {\bibfnamefont {G.~A.}\ \bibnamefont
  {Smith}}, \bibinfo {author} {\bibfnamefont {A.}~\bibnamefont {Silberfarb}},
  \bibinfo {author} {\bibfnamefont {I.~H.}\ \bibnamefont {Deutsch}}, \ and\
  \bibinfo {author} {\bibfnamefont {P.~S.}\ \bibnamefont {Jessen}},\
  }\href@noop {} {\bibfield  {journal} {\bibinfo  {journal} {Phys. Rev. Lett.}\
  }\textbf {\bibinfo {volume} {97}},\ \bibinfo {pages} {180403} (\bibinfo
  {year} {2006})}\BibitemShut {NoStop}%
\bibitem [{\citenamefont {Chase}\ and\ \citenamefont
  {Geremia}(2009)}]{ChaseGeremia09}%
  \BibitemOpen
  \bibfield  {author} {\bibinfo {author} {\bibfnamefont {B.}~\bibnamefont
  {Chase}}\ and\ \bibinfo {author} {\bibfnamefont {J.~M.}\ \bibnamefont
  {Geremia}},\ }\href@noop {} {\bibfield  {journal} {\bibinfo  {journal} {Phys.
  Rev. A}\ }\textbf {\bibinfo {volume} {79}},\ \bibinfo {pages} {023314}
  (\bibinfo {year} {2009})}\BibitemShut {NoStop}%
\bibitem [{\citenamefont {Steck}\ \emph {et~al.}(2006)\citenamefont {Steck},
  \citenamefont {Jacobs}, \citenamefont {Mabuchi}, \citenamefont {Habib},\ and\
  \citenamefont {Bhattacharya}}]{Steck.et.al.06}%
  \BibitemOpen
  \bibfield  {author} {\bibinfo {author} {\bibfnamefont {D.~A.}\ \bibnamefont
  {Steck}}, \bibinfo {author} {\bibfnamefont {K.}~\bibnamefont {Jacobs}},
  \bibinfo {author} {\bibfnamefont {H.}~\bibnamefont {Mabuchi}}, \bibinfo
  {author} {\bibfnamefont {S.}~\bibnamefont {Habib}}, \ and\ \bibinfo {author}
  {\bibnamefont {Bhattacharya}},\ }\href@noop {} {\bibfield  {journal}
  {\bibinfo  {journal} {Phys. Rev. A}\ }\textbf {\bibinfo {volume} {74}},\
  \bibinfo {pages} {012322} (\bibinfo {year} {2006})}\BibitemShut {NoStop}%
\bibitem [{\citenamefont {Shabani}\ and\ \citenamefont
  {Jacobs}(2008)}]{ShabaniJacobs08}%
  \BibitemOpen
  \bibfield  {author} {\bibinfo {author} {\bibfnamefont {A.}~\bibnamefont
  {Shabani}}\ and\ \bibinfo {author} {\bibfnamefont {K.}~\bibnamefont
  {Jacobs}},\ }\href@noop {} {\bibfield  {journal} {\bibinfo  {journal} {Phys.
  Rev. Lett.}\ }\textbf {\bibinfo {volume} {101}},\ \bibinfo {pages} {230403}
  (\bibinfo {year} {2008})}\BibitemShut {NoStop}%
\bibitem [{\citenamefont {Ruostekoski}\ and\ \citenamefont
  {Walls}(1997)}]{RW97}%
  \BibitemOpen
  \bibfield  {author} {\bibinfo {author} {\bibfnamefont {J.}~\bibnamefont
  {Ruostekoski}}\ and\ \bibinfo {author} {\bibfnamefont {D.~F.}\ \bibnamefont
  {Walls}},\ }\href {\doibase 10.1103/PhysRevA.56.2996} {\bibfield  {journal}
  {\bibinfo  {journal} {Phys. Rev. A}\ }\textbf {\bibinfo {volume} {56}},\
  \bibinfo {pages} {2996} (\bibinfo {year} {1997})}\BibitemShut {NoStop}%
\bibitem [{\citenamefont {Corney}\ and\ \citenamefont
  {Milburn}(1998)}]{CorneyMilburn98}%
  \BibitemOpen
  \bibfield  {author} {\bibinfo {author} {\bibfnamefont {J.}~\bibnamefont
  {Corney}}\ and\ \bibinfo {author} {\bibfnamefont {G.}~\bibnamefont
  {Milburn}},\ }\href@noop {} {\bibfield  {journal} {\bibinfo  {journal} {Phys.
  Rev. A}\ }\textbf {\bibinfo {volume} {58}},\ \bibinfo {pages} {2399}
  (\bibinfo {year} {1998})}\BibitemShut {NoStop}%
\bibitem [{\citenamefont {Dalvit}\ \emph {et~al.}(2002)\citenamefont {Dalvit},
  \citenamefont {Dziarmaga},\ and\ \citenamefont {Onofrio}}]{DDO02}%
  \BibitemOpen
  \bibfield  {author} {\bibinfo {author} {\bibfnamefont {D.~A.~R.}\
  \bibnamefont {Dalvit}}, \bibinfo {author} {\bibfnamefont {J.}~\bibnamefont
  {Dziarmaga}}, \ and\ \bibinfo {author} {\bibfnamefont {R.}~\bibnamefont
  {Onofrio}},\ }\href {\doibase 10.1103/PhysRevA.65.053604} {\bibfield
  {journal} {\bibinfo  {journal} {Physical Review A}\ }\textbf {\bibinfo
  {volume} {65}},\ \bibinfo {pages} {053604} (\bibinfo {year}
  {2002})}\BibitemShut {NoStop}%
\bibitem [{\citenamefont {Ruostekoski}\ and\ \citenamefont
  {Walls}(1998)}]{RW98}%
  \BibitemOpen
  \bibfield  {author} {\bibinfo {author} {\bibfnamefont {J.}~\bibnamefont
  {Ruostekoski}}\ and\ \bibinfo {author} {\bibfnamefont {D.~F.}\ \bibnamefont
  {Walls}},\ }\href {\doibase 10.1103/PhysRevA.58.R50} {\bibfield  {journal}
  {\bibinfo  {journal} {Phys. Rev. A}\ }\textbf {\bibinfo {volume} {58}},\
  \bibinfo {pages} {50(R)} (\bibinfo {year} {1998})}\BibitemShut {NoStop}%
\bibitem [{\citenamefont {Mekhov}\ \emph {et~al.}(2007)\citenamefont {Mekhov},
  \citenamefont {Maschler},\ and\ \citenamefont {Ritsch}}]{MMR07}%
  \BibitemOpen
  \bibfield  {author} {\bibinfo {author} {\bibfnamefont {I.~B.}\ \bibnamefont
  {Mekhov}}, \bibinfo {author} {\bibfnamefont {C.}~\bibnamefont {Maschler}}, \
  and\ \bibinfo {author} {\bibfnamefont {H.}~\bibnamefont {Ritsch}},\
  }\href@noop {} {\bibfield  {journal} {\bibinfo  {journal} {Phys. Rev. Lett.}\
  }\textbf {\bibinfo {volume} {98}},\ \bibinfo {pages} {100402} (\bibinfo
  {year} {2007})}\BibitemShut {NoStop}%
\bibitem [{\citenamefont {Mekhov}\ and\ \citenamefont {Ritsch}(2009)}]{MR09}%
  \BibitemOpen
  \bibfield  {author} {\bibinfo {author} {\bibfnamefont {I.~B.}\ \bibnamefont
  {Mekhov}}\ and\ \bibinfo {author} {\bibfnamefont {H.}~\bibnamefont
  {Ritsch}},\ }\href@noop {} {\bibfield  {journal} {\bibinfo  {journal} {Phys.
  Rev. Lett.}\ }\textbf {\bibinfo {volume} {102}},\ \bibinfo {pages} {020403}
  (\bibinfo {year} {2009})}\BibitemShut {NoStop}%
\bibitem [{\citenamefont {Li}\ and\ \citenamefont {Liu}(2006)}]{LL06}%
  \BibitemOpen
  \bibfield  {author} {\bibinfo {author} {\bibfnamefont {W.}~\bibnamefont
  {Li}}\ and\ \bibinfo {author} {\bibfnamefont {J.}~\bibnamefont {Liu}},\
  }\href {\doibase 10.1103/PhysRevA.74.063613} {\bibfield  {journal} {\bibinfo
  {journal} {Phys. Rev. A}\ }\textbf {\bibinfo {volume} {74}},\ \bibinfo
  {pages} {063613} (\bibinfo {year} {2006})}\BibitemShut {NoStop}%
\bibitem [{\citenamefont {Szigeti}\ \emph {et~al.}(2010)\citenamefont
  {Szigeti}, \citenamefont {Hush}, \citenamefont {Carvalho},\ and\
  \citenamefont {Hope}}]{SHC10}%
  \BibitemOpen
  \bibfield  {author} {\bibinfo {author} {\bibfnamefont {S.~S.}\ \bibnamefont
  {Szigeti}}, \bibinfo {author} {\bibfnamefont {M.~R.}\ \bibnamefont {Hush}},
  \bibinfo {author} {\bibfnamefont {A.~R.~R.}\ \bibnamefont {Carvalho}}, \ and\
  \bibinfo {author} {\bibfnamefont {J.~J.}\ \bibnamefont {Hope}},\ }\href
  {\doibase 10.1103/PhysRevA.82.043632} {\bibfield  {journal} {\bibinfo
  {journal} {Phys. Rev. A}\ }\textbf {\bibinfo {volume} {82}},\ \bibinfo
  {pages} {043632} (\bibinfo {year} {2010})}\BibitemShut {NoStop}%
\bibitem [{\citenamefont {Milburn}\ \emph {et~al.}(1997)\citenamefont
  {Milburn}, \citenamefont {Corney}, \citenamefont {Wright},\ and\
  \citenamefont {Walls}}]{MCWW97}%
  \BibitemOpen
  \bibfield  {author} {\bibinfo {author} {\bibfnamefont {G.~J.}\ \bibnamefont
  {Milburn}}, \bibinfo {author} {\bibfnamefont {J.}~\bibnamefont {Corney}},
  \bibinfo {author} {\bibfnamefont {E.~M.}\ \bibnamefont {Wright}}, \ and\
  \bibinfo {author} {\bibfnamefont {D.~F.}\ \bibnamefont {Walls}},\ }\href@noop
  {} {\bibfield  {journal} {\bibinfo  {journal} {Phys. Rev. A}\ }\textbf
  {\bibinfo {volume} {55}},\ \bibinfo {pages} {4318} (\bibinfo {year}
  {1997})}\BibitemShut {NoStop}%
\bibitem [{\citenamefont {Vardi}\ and\ \citenamefont {Anglin}(2000)}]{VA00}%
  \BibitemOpen
  \bibfield  {author} {\bibinfo {author} {\bibfnamefont {A.}~\bibnamefont
  {Vardi}}\ and\ \bibinfo {author} {\bibfnamefont {J.~R.}\ \bibnamefont
  {Anglin}},\ }\href@noop {} {\bibfield  {journal} {\bibinfo  {journal} {Phys.
  Rev. Lett.}\ }\textbf {\bibinfo {volume} {86}},\ \bibinfo {pages} {568}
  (\bibinfo {year} {2000})}\BibitemShut {NoStop}%
\bibitem [{\citenamefont {Zibold}\ \emph {et~al.}(2010)\citenamefont {Zibold},
  \citenamefont {Nicklas}, \citenamefont {Gross},\ and\ \citenamefont
  {Oberthaler}}]{ZNGO10}%
  \BibitemOpen
  \bibfield  {author} {\bibinfo {author} {\bibfnamefont {T.}~\bibnamefont
  {Zibold}}, \bibinfo {author} {\bibfnamefont {E.}~\bibnamefont {Nicklas}},
  \bibinfo {author} {\bibfnamefont {C.}~\bibnamefont {Gross}}, \ and\ \bibinfo
  {author} {\bibfnamefont {M.~K.}\ \bibnamefont {Oberthaler}},\ }\href@noop {}
  {\bibfield  {journal} {\bibinfo  {journal} {Phys. Rev. Lett.}\ }\textbf
  {\bibinfo {volume} {105}},\ \bibinfo {pages} {204101} (\bibinfo {year}
  {2010})}\BibitemShut {NoStop}%
\bibitem [{\citenamefont {Diosi}\ \emph {et~al.}(2006)\citenamefont {Diosi},
  \citenamefont {Konrad}, \citenamefont {Scherer},\ and\ \citenamefont
  {Audretsch}}]{Diosi.et.al06}%
  \BibitemOpen
  \bibfield  {author} {\bibinfo {author} {\bibfnamefont {L.}~\bibnamefont
  {Diosi}}, \bibinfo {author} {\bibfnamefont {T.}~\bibnamefont {Konrad}},
  \bibinfo {author} {\bibfnamefont {A.}~\bibnamefont {Scherer}}, \ and\
  \bibinfo {author} {\bibfnamefont {J.}~\bibnamefont {Audretsch}},\ }\href@noop
  {} {\bibfield  {journal} {\bibinfo  {journal} {J. Phys. A}\ }\textbf
  {\bibinfo {volume} {39}},\ \bibinfo {pages} {L575} (\bibinfo {year}
  {2006})}\BibitemShut {NoStop}%
\bibitem [{\citenamefont {Konrad}\ \emph {et~al.}(2010)\citenamefont {Konrad},
  \citenamefont {Rothe}, \citenamefont {Petruccione},\ and\ \citenamefont
  {Diosi}}]{Konrad.et.al10}%
  \BibitemOpen
  \bibfield  {author} {\bibinfo {author} {\bibfnamefont {T.}~\bibnamefont
  {Konrad}}, \bibinfo {author} {\bibfnamefont {A.}~\bibnamefont {Rothe}},
  \bibinfo {author} {\bibfnamefont {F.}~\bibnamefont {Petruccione}}, \ and\
  \bibinfo {author} {\bibfnamefont {L.}~\bibnamefont {Diosi}},\ }\href@noop {}
  {\bibfield  {journal} {\bibinfo  {journal} {New J. Phys.}\ }\textbf {\bibinfo
  {volume} {12}},\ \bibinfo {pages} {043038} (\bibinfo {year}
  {2010})}\BibitemShut {NoStop}%
\bibitem [{\citenamefont {T.Konrad}\ and\ \citenamefont
  {H.Uys}(2012)}]{KonradUys12}%
  \BibitemOpen
  \bibfield  {author} {\bibinfo {author} {\bibnamefont {T.Konrad}}\ and\
  \bibinfo {author} {\bibnamefont {H.Uys}},\ }\href@noop {} {\bibfield
  {journal} {\bibinfo  {journal} {Phys. Rev. A}\ }\textbf {\bibinfo {volume}
  {85}},\ \bibinfo {pages} {012102} (\bibinfo {year} {2012})}\BibitemShut
  {NoStop}%
\bibitem [{\citenamefont {Jaksch}\ \emph {et~al.}(1998)\citenamefont {Jaksch},
  \citenamefont {Bruder}, \citenamefont {Cirac}, \citenamefont {Gardiner},\
  and\ \citenamefont {Zoller}}]{JBCGZ98}%
  \BibitemOpen
  \bibfield  {author} {\bibinfo {author} {\bibfnamefont {D.}~\bibnamefont
  {Jaksch}}, \bibinfo {author} {\bibfnamefont {C.}~\bibnamefont {Bruder}},
  \bibinfo {author} {\bibfnamefont {J.~I.}\ \bibnamefont {Cirac}}, \bibinfo
  {author} {\bibfnamefont {C.~W.}\ \bibnamefont {Gardiner}}, \ and\ \bibinfo
  {author} {\bibfnamefont {P.}~\bibnamefont {Zoller}},\ }\href@noop {}
  {\bibfield  {journal} {\bibinfo  {journal} {Phys. Rev. Lett.}\ }\textbf
  {\bibinfo {volume} {81}},\ \bibinfo {pages} {3108} (\bibinfo {year}
  {1998})}\BibitemShut {NoStop}%
\bibitem [{Note1()}]{Note1}%
  \BibitemOpen
  \bibinfo {note} {\label {foot.bias} Numerically, we add a small bias of the
  order of $10^{-2}\protect \hat n_1$, in order to avoid unstable macroscopic
  superposition states \cite {DHC07} that result in numerical
  artifacts.}\BibitemShut {Stop}%
\bibitem [{\citenamefont {Morsch}\ and\ \citenamefont
  {Oberthaler}(2006)}]{MO06}%
  \BibitemOpen
  \bibfield  {author} {\bibinfo {author} {\bibfnamefont {O.}~\bibnamefont
  {Morsch}}\ and\ \bibinfo {author} {\bibfnamefont {M.}~\bibnamefont
  {Oberthaler}},\ }\href@noop {} {\bibfield  {journal} {\bibinfo  {journal}
  {Rev. Mod. Phys.}\ }\textbf {\bibinfo {volume} {78}},\ \bibinfo {pages} {179}
  (\bibinfo {year} {2006})}\BibitemShut {NoStop}%
\bibitem [{\citenamefont {Chin}\ \emph {et~al.}(2010)\citenamefont {Chin},
  \citenamefont {Grimm}, \citenamefont {Julienne},\ and\ \citenamefont
  {Tiesinga}}]{CGJT10}%
  \BibitemOpen
  \bibfield  {author} {\bibinfo {author} {\bibfnamefont {C.}~\bibnamefont
  {Chin}}, \bibinfo {author} {\bibfnamefont {R.}~\bibnamefont {Grimm}},
  \bibinfo {author} {\bibfnamefont {P.}~\bibnamefont {Julienne}}, \ and\
  \bibinfo {author} {\bibfnamefont {E.}~\bibnamefont {Tiesinga}},\ }\href
  {\doibase 10.1103/RevModPhys.82.1225} {\bibfield  {journal} {\bibinfo
  {journal} {Rev. Mod. Phys.}\ }\textbf {\bibinfo {volume} {82}},\ \bibinfo
  {pages} {1225} (\bibinfo {year} {2010})}\BibitemShut {NoStop}%
\bibitem [{\citenamefont {Chuchem}\ \emph {et~al.}(2010)\citenamefont
  {Chuchem}, \citenamefont {Smith-Mannschott}, \citenamefont {Hiller},
  \citenamefont {Kottos}, \citenamefont {Vardi},\ and\ \citenamefont
  {Cohen}}]{Chuc10}%
  \BibitemOpen
  \bibfield  {author} {\bibinfo {author} {\bibfnamefont {M.}~\bibnamefont
  {Chuchem}}, \bibinfo {author} {\bibfnamefont {K.}~\bibnamefont
  {Smith-Mannschott}}, \bibinfo {author} {\bibfnamefont {M.}~\bibnamefont
  {Hiller}}, \bibinfo {author} {\bibfnamefont {T.}~\bibnamefont {Kottos}},
  \bibinfo {author} {\bibfnamefont {A.}~\bibnamefont {Vardi}}, \ and\ \bibinfo
  {author} {\bibfnamefont {D.}~\bibnamefont {Cohen}},\ }\href {\doibase
  10.1103/PhysRevA.82.053617} {\bibfield  {journal} {\bibinfo  {journal} {Phys.
  Rev. A}\ }\textbf {\bibinfo {volume} {82}},\ \bibinfo {pages} {053617}
  (\bibinfo {year} {2010})}\BibitemShut {NoStop}%
\bibitem [{\citenamefont {Bernstein}\ \emph {et~al.}(1990)\citenamefont
  {Bernstein}, \citenamefont {Eilbeck},\ and\ \citenamefont {Scott}}]{BES90}%
  \BibitemOpen
  \bibfield  {author} {\bibinfo {author} {\bibfnamefont {L.}~\bibnamefont
  {Bernstein}}, \bibinfo {author} {\bibfnamefont {J.~C.}\ \bibnamefont
  {Eilbeck}}, \ and\ \bibinfo {author} {\bibfnamefont {A.~C.}\ \bibnamefont
  {Scott}},\ }\href@noop {} {\bibfield  {journal} {\bibinfo  {journal}
  {Nonlinearity}\ }\textbf {\bibinfo {volume} {3}},\ \bibinfo {pages} {293}
  (\bibinfo {year} {1990})}\BibitemShut {NoStop}%
\bibitem [{\citenamefont {Smerzi}\ \emph {et~al.}(1997)\citenamefont {Smerzi},
  \citenamefont {Fantoni}, \citenamefont {Giovanazzi},\ and\ \citenamefont
  {Shenoy}}]{SFGS97}%
  \BibitemOpen
  \bibfield  {author} {\bibinfo {author} {\bibfnamefont {A.}~\bibnamefont
  {Smerzi}}, \bibinfo {author} {\bibfnamefont {S.}~\bibnamefont {Fantoni}},
  \bibinfo {author} {\bibfnamefont {S.}~\bibnamefont {Giovanazzi}}, \ and\
  \bibinfo {author} {\bibfnamefont {S.}~\bibnamefont {Shenoy}},\ }\href@noop {}
  {\bibfield  {journal} {\bibinfo  {journal} {Phys. Rev. Lett.}\ }\textbf
  {\bibinfo {volume} {79}},\ \bibinfo {pages} {4950} (\bibinfo {year}
  {1997})}\BibitemShut {NoStop}%
\bibitem [{\citenamefont {Albiez}\ \emph {et~al.}(2005)\citenamefont {Albiez},
  \citenamefont {Gati}, \citenamefont {Folling}, \citenamefont {Hunsmann},
  \citenamefont {Cristiani},\ and\ \citenamefont {Oberthaler}}]{AGFHCO05}%
  \BibitemOpen
  \bibfield  {author} {\bibinfo {author} {\bibfnamefont {M.}~\bibnamefont
  {Albiez}}, \bibinfo {author} {\bibfnamefont {R.}~\bibnamefont {Gati}},
  \bibinfo {author} {\bibfnamefont {J.}~\bibnamefont {Folling}}, \bibinfo
  {author} {\bibfnamefont {S.}~\bibnamefont {Hunsmann}}, \bibinfo {author}
  {\bibfnamefont {M.}~\bibnamefont {Cristiani}}, \ and\ \bibinfo {author}
  {\bibfnamefont {M.~K.}\ \bibnamefont {Oberthaler}},\ }\href@noop {}
  {\bibfield  {journal} {\bibinfo  {journal} {Phys. Rev. Lett.}\ }\textbf
  {\bibinfo {volume} {95}},\ \bibinfo {pages} {010402} (\bibinfo {year}
  {2005})}\BibitemShut {NoStop}%
\bibitem [{\citenamefont {Javanainen}\ and\ \citenamefont
  {Ruostekoski}(2011)}]{JJ11}%
  \BibitemOpen
  \bibfield  {author} {\bibinfo {author} {\bibfnamefont {J.}~\bibnamefont
  {Javanainen}}\ and\ \bibinfo {author} {\bibfnamefont {J.}~\bibnamefont
  {Ruostekoski}},\ }\href@noop {} {\bibfield  {journal} {\bibinfo  {journal}
  {arxiv}\ ,\ \bibinfo {pages} {1104.0820}} (\bibinfo {year}
  {2011})}\BibitemShut {NoStop}%
\bibitem [{\citenamefont {Sanders}\ \emph {et~al.}(2010)\citenamefont
  {Sanders}, \citenamefont {Mintert},\ and\ \citenamefont {Heller}}]{SMH10}%
  \BibitemOpen
  \bibfield  {author} {\bibinfo {author} {\bibfnamefont {S.~N.}\ \bibnamefont
  {Sanders}}, \bibinfo {author} {\bibfnamefont {F.}~\bibnamefont {Mintert}}, \
  and\ \bibinfo {author} {\bibfnamefont {E.~J.}\ \bibnamefont {Heller}},\
  }\href@noop {} {\bibfield  {journal} {\bibinfo  {journal} {Phys. Rev. Lett.}\
  }\textbf {\bibinfo {volume} {105}},\ \bibinfo {pages} {035301} (\bibinfo
  {year} {2010})}\BibitemShut {NoStop}%
\bibitem [{\citenamefont {Hunn}\ \emph {et~al.}(2011)\citenamefont {Hunn},
  \citenamefont {Hiller}, \citenamefont {Buchleitner}, \citenamefont {Cohen},\
  and\ \citenamefont {Kottos}}]{SOD11}%
  \BibitemOpen
  \bibfield  {author} {\bibinfo {author} {\bibfnamefont {S.}~\bibnamefont
  {Hunn}}, \bibinfo {author} {\bibfnamefont {M.}~\bibnamefont {Hiller}},
  \bibinfo {author} {\bibfnamefont {A.}~\bibnamefont {Buchleitner}}, \bibinfo
  {author} {\bibfnamefont {D.}~\bibnamefont {Cohen}}, \ and\ \bibinfo {author}
  {\bibfnamefont {T.}~\bibnamefont {Kottos}},\ }\href@noop {} {\bibfield
  {journal} {\bibinfo  {journal} {EPJ D}\ }\textbf {\bibinfo {volume} {63}},\
  \bibinfo {pages} {55} (\bibinfo {year} {2011})}\BibitemShut {NoStop}%
\bibitem [{\citenamefont {Hunn}\ \emph {et~al.}(2012)\citenamefont {Hunn},
  \citenamefont {Hiller}, \citenamefont {Cohen}, \citenamefont {Kottos},\ and\
  \citenamefont {Buchleitner}}]{Hunn12}%
  \BibitemOpen
  \bibfield  {author} {\bibinfo {author} {\bibfnamefont {S.}~\bibnamefont
  {Hunn}}, \bibinfo {author} {\bibfnamefont {M.}~\bibnamefont {Hiller}},
  \bibinfo {author} {\bibfnamefont {D.}~\bibnamefont {Cohen}}, \bibinfo
  {author} {\bibfnamefont {T.}~\bibnamefont {Kottos}}, \ and\ \bibinfo {author}
  {\bibfnamefont {A.}~\bibnamefont {Buchleitner}},\ }\href {\doibase
  10.1088/0953-4075/45/8/085302} {\bibfield  {journal} {\bibinfo  {journal} {J.
  Phys. B}\ }\textbf {\bibinfo {volume} {45}},\ \bibinfo {pages} {085302}
  (\bibinfo {year} {2012})}\BibitemShut {NoStop}%
\bibitem [{\citenamefont {Milburn}\ \emph {et~al.}(1994)\citenamefont
  {Milburn}, \citenamefont {Jacobs},\ and\ \citenamefont
  {Walls}}]{Milburn.et.al94}%
  \BibitemOpen
  \bibfield  {author} {\bibinfo {author} {\bibfnamefont {G.}~\bibnamefont
  {Milburn}}, \bibinfo {author} {\bibfnamefont {K.}~\bibnamefont {Jacobs}}, \
  and\ \bibinfo {author} {\bibfnamefont {D.}~\bibnamefont {Walls}},\
  }\href@noop {} {\bibfield  {journal} {\bibinfo  {journal} {Phys. Rev. A}\
  }\textbf {\bibinfo {volume} {50}},\ \bibinfo {pages} {5256} (\bibinfo {year}
  {1994})}\BibitemShut {NoStop}%
\bibitem [{\citenamefont {Di{\'o}si}(1988)}]{Diosi88}%
  \BibitemOpen
  \bibfield  {author} {\bibinfo {author} {\bibfnamefont {L.}~\bibnamefont
  {Di{\'o}si}},\ }\href@noop {} {\bibfield  {journal} {\bibinfo  {journal}
  {Phys. Lett. A}\ }\textbf {\bibinfo {volume} {129}},\ \bibinfo {pages} {419}
  (\bibinfo {year} {1988})}\BibitemShut {NoStop}%
\bibitem [{\citenamefont {Sondermann}(1998)}]{Sondermann98}%
  \BibitemOpen
  \bibfield  {author} {\bibinfo {author} {\bibfnamefont {D.}~\bibnamefont
  {Sondermann}},\ }\emph {\bibinfo {title} {Die Iteration nichtzerst{\"o}render
  quantenmechanischer {M}essungen}},\ \href@noop {} {Ph.D. thesis},\ \bibinfo
  {school} {University of G\"ottingen}, \bibinfo {address} {Germany} (\bibinfo
  {year} {1998})\BibitemShut {NoStop}%
\bibitem [{\citenamefont {Audretsch}\ \emph
  {et~al.}(2002{\natexlab{b}})\citenamefont {Audretsch}, \citenamefont
  {Di{\'o}si},\ and\ \citenamefont {Konrad}}]{AudretschDiosiKonrad02}%
  \BibitemOpen
  \bibfield  {author} {\bibinfo {author} {\bibfnamefont {J.}~\bibnamefont
  {Audretsch}}, \bibinfo {author} {\bibfnamefont {L.}~\bibnamefont
  {Di{\'o}si}}, \ and\ \bibinfo {author} {\bibfnamefont {T.}~\bibnamefont
  {Konrad}},\ }\href@noop {} {\bibfield  {journal} {\bibinfo  {journal} {Phys.
  Rev. A}\ }\textbf {\bibinfo {volume} {66}},\ \bibinfo {pages} {022310}
  (\bibinfo {year} {2002}{\natexlab{b}})}\BibitemShut {NoStop}%
\bibitem [{Note2()}]{Note2}%
  \BibitemOpen
  \bibinfo {note} {The limit of $\protect \bar {\gamma } \rightarrow \infty $
  corresponds to the Zeno regime, where the dynamics is brought to a
  halt.}\BibitemShut {Stop}%
\bibitem [{\citenamefont {Khodorkovsky}\ \emph {et~al.}(2008)\citenamefont
  {Khodorkovsky}, \citenamefont {Kurizki},\ and\ \citenamefont
  {Vardi}}]{KKV08}%
  \BibitemOpen
  \bibfield  {author} {\bibinfo {author} {\bibfnamefont {Y.}~\bibnamefont
  {Khodorkovsky}}, \bibinfo {author} {\bibfnamefont {G.}~\bibnamefont
  {Kurizki}}, \ and\ \bibinfo {author} {\bibfnamefont {A.}~\bibnamefont
  {Vardi}},\ }\href {\doibase 10.1103/PhysRevLett.100.220403} {\bibfield
  {journal} {\bibinfo  {journal} {Phys. Rev. Lett.}\ }\textbf {\bibinfo
  {volume} {100}},\ \bibinfo {pages} {220403} (\bibinfo {year}
  {2008})}\BibitemShut {NoStop}%
\bibitem [{\citenamefont {Ferrini}\ \emph {et~al.}(2010)\citenamefont
  {Ferrini}, \citenamefont {Spehner}, \citenamefont {Minguzzi},\ and\
  \citenamefont {Hekking}}]{FSM10}%
  \BibitemOpen
  \bibfield  {author} {\bibinfo {author} {\bibfnamefont {G.}~\bibnamefont
  {Ferrini}}, \bibinfo {author} {\bibfnamefont {D.}~\bibnamefont {Spehner}},
  \bibinfo {author} {\bibfnamefont {A.}~\bibnamefont {Minguzzi}}, \ and\
  \bibinfo {author} {\bibfnamefont {F.~W.~J.}\ \bibnamefont {Hekking}},\ }\href
  {\doibase 10.1103/PhysRevA.82.033621} {\bibfield  {journal} {\bibinfo
  {journal} {Phys. Rev. A}\ }\textbf {\bibinfo {volume} {82}},\ \bibinfo
  {pages} {033621} (\bibinfo {year} {2010})}\BibitemShut {NoStop}%
\bibitem [{\citenamefont {{Bar-Gill}}\ \emph {et~al.}(2009)\citenamefont
  {{Bar-Gill}}, \citenamefont {Kurizki}, \citenamefont {Oberthaler},\ and\
  \citenamefont {Davidson}}]{BKOD09}%
  \BibitemOpen
  \bibfield  {author} {\bibinfo {author} {\bibfnamefont {N.}~\bibnamefont
  {{Bar-Gill}}}, \bibinfo {author} {\bibfnamefont {G.}~\bibnamefont {Kurizki}},
  \bibinfo {author} {\bibfnamefont {M.}~\bibnamefont {Oberthaler}}, \ and\
  \bibinfo {author} {\bibfnamefont {N.}~\bibnamefont {Davidson}},\ }\href
  {\doibase 10.1103/PhysRevA.80.053613} {\bibfield  {journal} {\bibinfo
  {journal} {Phys. Rev. A}\ }\textbf {\bibinfo {volume} {80}},\ \bibinfo
  {pages} {053613} (\bibinfo {year} {2009})}\BibitemShut {NoStop}%
\bibitem [{\citenamefont {Khripkov}\ \emph {et~al.}(2012)\citenamefont
  {Khripkov}, \citenamefont {Vardi},\ and\ \citenamefont {Cohen}}]{KVC12}%
  \BibitemOpen
  \bibfield  {author} {\bibinfo {author} {\bibfnamefont {C.}~\bibnamefont
  {Khripkov}}, \bibinfo {author} {\bibfnamefont {A.}~\bibnamefont {Vardi}}, \
  and\ \bibinfo {author} {\bibfnamefont {D.}~\bibnamefont {Cohen}},\
  }\href@noop {} {\bibfield  {journal} {\bibinfo  {journal} {Phys. Rev. A}\
  }\textbf {\bibinfo {volume} {85}},\ \bibinfo {pages} {053632} (\bibinfo
  {year} {2012})}\BibitemShut {NoStop}%
\bibitem [{\citenamefont {Press}\ \emph {et~al.}(1999)\citenamefont {Press},
  \citenamefont {Teukolsky}, \citenamefont {Vetterling},\ and\ \citenamefont
  {Flannery}}]{Press.et.al99}%
  \BibitemOpen
  \bibfield  {author} {\bibinfo {author} {\bibfnamefont {W.}~\bibnamefont
  {Press}}, \bibinfo {author} {\bibfnamefont {S.}~\bibnamefont {Teukolsky}},
  \bibinfo {author} {\bibfnamefont {W.}~\bibnamefont {Vetterling}}, \ and\
  \bibinfo {author} {\bibfnamefont {B.}~\bibnamefont {Flannery}},\ }\href@noop
  {} {\emph {\bibinfo {title} {Numerical Recipes in C}}}\ (\bibinfo
  {publisher} {Cambridge University Press},\ \bibinfo {year}
  {1999})\BibitemShut {NoStop}%
\bibitem [{\citenamefont {Schmied}\ and\ \citenamefont
  {Treutlein}(2011)}]{ST11}%
  \BibitemOpen
  \bibfield  {author} {\bibinfo {author} {\bibfnamefont {R.}~\bibnamefont
  {Schmied}}\ and\ \bibinfo {author} {\bibfnamefont {P.}~\bibnamefont
  {Treutlein}},\ }\href {\doibase 10.1088/1367-2630/13/6/065019} {\bibfield
  {journal} {\bibinfo  {journal} {New J. Phys.}\ }\textbf {\bibinfo {volume}
  {13}},\ \bibinfo {pages} {065019} (\bibinfo {year} {2011})}\BibitemShut
  {NoStop}%
\bibitem [{\citenamefont {Christandl}\ and\ \citenamefont
  {Renner}(2011)}]{CR11}%
  \BibitemOpen
  \bibfield  {author} {\bibinfo {author} {\bibfnamefont {M.}~\bibnamefont
  {Christandl}}\ and\ \bibinfo {author} {\bibfnamefont {R.}~\bibnamefont
  {Renner}},\ }\href {http://arxiv.org/abs/1108.5329} {\bibfield  {journal}
  {\bibinfo  {journal} {{arXiv:1108.5329}}\ } (\bibinfo {year}
  {2011})}\BibitemShut {NoStop}%
\bibitem [{\citenamefont {Doherty}\ and\ \citenamefont
  {Jacobs}(1999)}]{DohertyJacobs99}%
  \BibitemOpen
  \bibfield  {author} {\bibinfo {author} {\bibfnamefont {A.}~\bibnamefont
  {Doherty}}\ and\ \bibinfo {author} {\bibfnamefont {K.}~\bibnamefont
  {Jacobs}},\ }\href@noop {} {\bibfield  {journal} {\bibinfo  {journal} {Phys.
  Rev. A}\ }\textbf {\bibinfo {volume} {60}},\ \bibinfo {pages} {2700}
  (\bibinfo {year} {1999})}\BibitemShut {NoStop}%
\bibitem [{\citenamefont {Doherty}\ \emph {et~al.}(1999)\citenamefont
  {Doherty}, \citenamefont {Tan}, \citenamefont {Parkins},\ and\ \citenamefont
  {Walls}}]{Doherty.et.al00}%
  \BibitemOpen
  \bibfield  {author} {\bibinfo {author} {\bibfnamefont {A.}~\bibnamefont
  {Doherty}}, \bibinfo {author} {\bibfnamefont {S.}~\bibnamefont {Tan}},
  \bibinfo {author} {\bibfnamefont {A.}~\bibnamefont {Parkins}}, \ and\
  \bibinfo {author} {\bibfnamefont {D.}~\bibnamefont {Walls}},\ }\href@noop {}
  {\bibfield  {journal} {\bibinfo  {journal} {Phys.\ Rev.}\ }\textbf {\bibinfo
  {volume} {A 60}},\ \bibinfo {pages} {2380} (\bibinfo {year}
  {1999})}\BibitemShut {NoStop}%
\bibitem [{\citenamefont {Rouchon}(2011)}]{Rouchon11}%
  \BibitemOpen
  \bibfield  {author} {\bibinfo {author} {\bibfnamefont {P.}~\bibnamefont
  {Rouchon}},\ }\href@noop {} {\bibfield  {journal} {\bibinfo  {journal} {IEEE
  Transactions on automatic control}\ }\textbf {\bibinfo {volume} {56}},\
  \bibinfo {pages} {2743} (\bibinfo {year} {2011})}\BibitemShut {NoStop}%
\bibitem [{\citenamefont {Amini}\ \emph {et~al.}(2011)\citenamefont {Amini},
  \citenamefont {Mirrahimi},\ and\ \citenamefont
  {Rouchon}}]{AminiMazyarRouchon11}%
  \BibitemOpen
  \bibfield  {author} {\bibinfo {author} {\bibfnamefont {H.}~\bibnamefont
  {Amini}}, \bibinfo {author} {\bibfnamefont {M.}~\bibnamefont {Mirrahimi}}, \
  and\ \bibinfo {author} {\bibfnamefont {P.}~\bibnamefont {Rouchon}},\ }in\
  \href@noop {} {\emph {\bibinfo {booktitle} {Decision and Control and European
  Control Conference}}}\ (\bibinfo {year} {2011})\ p.\ \bibinfo {pages}
  {6242}\BibitemShut {NoStop}%
\bibitem [{Note4()}]{Note4}%
  \BibitemOpen
  \bibinfo {note} {Animated movies of the (un-)monitored evolution for $u=0$
  and $u=1$ can be found in the supplementary online material.}\BibitemShut
  {Stop}%
\bibitem [{\citenamefont {Zhang}\ \emph {et~al.}(1990)\citenamefont {Zhang},
  \citenamefont {Feng},\ and\ \citenamefont {Gilmore}}]{ZFG90}%
  \BibitemOpen
  \bibfield  {author} {\bibinfo {author} {\bibfnamefont {W.-M.}\ \bibnamefont
  {Zhang}}, \bibinfo {author} {\bibfnamefont {D.~H.}\ \bibnamefont {Feng}}, \
  and\ \bibinfo {author} {\bibfnamefont {R.}~\bibnamefont {Gilmore}},\ }\href
  {\doibase 10.1103/RevModPhys.62.867} {\bibfield  {journal} {\bibinfo
  {journal} {Rev. Mod. Phys.}\ }\textbf {\bibinfo {volume} {62}},\ \bibinfo
  {pages} {867} (\bibinfo {year} {1990})}\BibitemShut {NoStop}%
\bibitem [{Note3()}]{Note3}%
  \BibitemOpen
  \bibinfo {note} {Together with the state $|-j\protect \rangle $ which
  represents a preparation on the South pole of the Bloch sphere, these are the
  only two Fock-states which are also coherent states, similar to the vacuum
  state of a harmonic oscillator.}\BibitemShut {Stop}%
\bibitem [{Note5()}]{Note5}%
  \BibitemOpen
  \bibinfo {note} {The nonlinear rotation is manifest in the ``curved''
  mean-field trajectories for $u=1$ in contrast to the ``straight''
  trajectories that result from plane-sphere intersections for $u=0$ (cp. thin
  black lines in Fig.~\ref {fig.results_u0}a) and Fig.~\ref
  {fig.results_u1}a))}\BibitemShut {NoStop}%
\bibitem [{Note6()}]{Note6}%
  \BibitemOpen
  \bibinfo {note} {A detailed discussion of the asymptotic behavior of the
  Bloch vector for various initial conditions and interaction values can be
  found in \cite {Chuc10}.}\BibitemShut {Stop}%
\bibitem [{Note7()}]{Note7}%
  \BibitemOpen
  \bibinfo {note} {We found that $\protect \bar {\gamma }$ should be greater or
  equal to $u$ to prevent the damping of the Rabi oscillations but of the order
  of $K$ in order to not significantly disturb them. A quantitative analysis of
  the minimal measurement strength should be based on a comparison of the
  dephasing rate (specific to the state under investigation at a given value of
  $u$) and the measurement strength $\protect \bar {\gamma }$.}\BibitemShut
  {Stop}%
\bibitem [{\citenamefont {Busch}\ \emph {et~al.}(1995)\citenamefont {Busch},
  \citenamefont {Grabowski},\ and\ \citenamefont
  {Lahti}}]{BuschGrabowskiLahti95}%
  \BibitemOpen
  \bibfield  {author} {\bibinfo {author} {\bibfnamefont {P.}~\bibnamefont
  {Busch}}, \bibinfo {author} {\bibfnamefont {M.}~\bibnamefont {Grabowski}}, \
  and\ \bibinfo {author} {\bibfnamefont {J.}~\bibnamefont {Lahti}},\
  }\href@noop {} {\emph {\bibinfo {title} {Operational Quantum Physics}}}\
  (\bibinfo  {publisher} {Springer Verlag},\ \bibinfo {address} {Heidelberg},\
  \bibinfo {year} {1995})\BibitemShut {NoStop}%
\bibitem [{Note8()}]{Note8}%
  \BibitemOpen
  \bibinfo {note} {See \cite {ST11,CR11} and references therein, for recent
  advances in the tomography of a BEC in a double-well potential}\BibitemShut
  {NoStop}%
\bibitem [{\citenamefont {Eckardt}\ \emph {et~al.}(2009)\citenamefont
  {Eckardt}, \citenamefont {Holthaus}, \citenamefont {Lignier}, \citenamefont
  {Zenesini}, \citenamefont {Ciampini}, \citenamefont {Morsch},\ and\
  \citenamefont {Arimondo}}]{Ecke_etal09}%
  \BibitemOpen
  \bibfield  {author} {\bibinfo {author} {\bibfnamefont {A.}~\bibnamefont
  {Eckardt}}, \bibinfo {author} {\bibfnamefont {M.}~\bibnamefont {Holthaus}},
  \bibinfo {author} {\bibfnamefont {H.}~\bibnamefont {Lignier}}, \bibinfo
  {author} {\bibfnamefont {A.}~\bibnamefont {Zenesini}}, \bibinfo {author}
  {\bibfnamefont {D.}~\bibnamefont {Ciampini}}, \bibinfo {author}
  {\bibfnamefont {O.}~\bibnamefont {Morsch}}, \ and\ \bibinfo {author}
  {\bibfnamefont {E.}~\bibnamefont {Arimondo}},\ }\href@noop {} {\bibfield
  {journal} {\bibinfo  {journal} {Phys. Rev. A}\ }\textbf {\bibinfo {volume}
  {79}},\ \bibinfo {pages} {013611} (\bibinfo {year} {2009})}\BibitemShut
  {NoStop}%
\bibitem [{\citenamefont {Hanson}\ and\ \citenamefont
  {Lambropoulos}(1995)}]{HL95}%
  \BibitemOpen
  \bibfield  {author} {\bibinfo {author} {\bibfnamefont {L.~G.}\ \bibnamefont
  {Hanson}}\ and\ \bibinfo {author} {\bibfnamefont {P.}~\bibnamefont
  {Lambropoulos}},\ }\href {\doibase 10.1103/PhysRevLett.74.5009} {\bibfield
  {journal} {\bibinfo  {journal} {Phys. Rev. Lett.}\ }\textbf {\bibinfo
  {volume} {74}},\ \bibinfo {pages} {5009} (\bibinfo {year}
  {1995})}\BibitemShut {NoStop}%
\bibitem [{\citenamefont {Buchleitner}\ \emph {et~al.}(2002)\citenamefont
  {Buchleitner}, \citenamefont {Delande},\ and\ \citenamefont
  {Zakrzewski}}]{BDZ02}%
  \BibitemOpen
  \bibfield  {author} {\bibinfo {author} {\bibfnamefont {A.}~\bibnamefont
  {Buchleitner}}, \bibinfo {author} {\bibfnamefont {D.}~\bibnamefont
  {Delande}}, \ and\ \bibinfo {author} {\bibfnamefont {J.}~\bibnamefont
  {Zakrzewski}},\ }\href {\doibase PII S0370-1573(02)00270-3} {\bibfield
  {journal} {\bibinfo  {journal} {Phys. Rep.}\ }\textbf {\bibinfo {volume}
  {368}},\ \bibinfo {pages} {409} (\bibinfo {year} {2002})}\BibitemShut
  {NoStop}%
\bibitem [{\citenamefont {Dunning}\ \emph {et~al.}(2009)\citenamefont
  {Dunning}, \citenamefont {Mestayer}, \citenamefont {Reinhold}, \citenamefont
  {Yoshida},\ and\ \citenamefont {Burgd{\"o}rfer}}]{DMR09}%
  \BibitemOpen
  \bibfield  {author} {\bibinfo {author} {\bibfnamefont {F.~B.}\ \bibnamefont
  {Dunning}}, \bibinfo {author} {\bibfnamefont {J.~J.}\ \bibnamefont
  {Mestayer}}, \bibinfo {author} {\bibfnamefont {C.~O.}\ \bibnamefont
  {Reinhold}}, \bibinfo {author} {\bibfnamefont {S.}~\bibnamefont {Yoshida}}, \
  and\ \bibinfo {author} {\bibfnamefont {J.}~\bibnamefont {Burgd{\"o}rfer}},\
  }\href {\doibase 10.1088/0953-4075/42/2/022001} {\bibfield  {journal}
  {\bibinfo  {journal} {J. Phys. B}\ }\textbf {\bibinfo {volume} {42}},\
  \bibinfo {pages} {022001} (\bibinfo {year} {2009})}\BibitemShut {NoStop}%
\bibitem [{\citenamefont {Gross}\ \emph {et~al.}(2010)\citenamefont {Gross},
  \citenamefont {Zibold}, \citenamefont {Nicklas}, \citenamefont {Est{\`e}ve},\
  and\ \citenamefont {Oberthaler}}]{Gross_etal10}%
  \BibitemOpen
  \bibfield  {author} {\bibinfo {author} {\bibfnamefont {C.}~\bibnamefont
  {Gross}}, \bibinfo {author} {\bibfnamefont {T.}~\bibnamefont {Zibold}},
  \bibinfo {author} {\bibfnamefont {E.}~\bibnamefont {Nicklas}}, \bibinfo
  {author} {\bibfnamefont {J.}~\bibnamefont {Est{\`e}ve}}, \ and\ \bibinfo
  {author} {\bibfnamefont {M.~K.}\ \bibnamefont {Oberthaler}},\ }\href@noop {}
  {\bibfield  {journal} {\bibinfo  {journal} {Nature}\ }\textbf {\bibinfo
  {volume} {464}},\ \bibinfo {pages} {1165} (\bibinfo {year}
  {2010})}\BibitemShut {NoStop}%
\bibitem [{\citenamefont {Dounas-Frazer}\ \emph {et~al.}(2007)\citenamefont
  {Dounas-Frazer}, \citenamefont {Hermundstad},\ and\ \citenamefont
  {Carr}}]{DHC07}%
  \BibitemOpen
  \bibfield  {author} {\bibinfo {author} {\bibfnamefont {D.~R.}\ \bibnamefont
  {Dounas-Frazer}}, \bibinfo {author} {\bibfnamefont {A.~M.}\ \bibnamefont
  {Hermundstad}}, \ and\ \bibinfo {author} {\bibfnamefont {L.~D.}\ \bibnamefont
  {Carr}},\ }\href@noop {} {\bibfield  {journal} {\bibinfo  {journal} {Phys.
  Rev. Lett.}\ }\textbf {\bibinfo {volume} {99}},\ \bibinfo {pages} {200402}
  (\bibinfo {year} {2007})}\BibitemShut {NoStop}%
\end{thebibliography}%
\end{document}